\documentclass[aps,superscriptaddress,preprint,amsmath,amssymb,nofootinbib,]{revtex4-1}

\usepackage{graphicx}
\usepackage{dcolumn}
\usepackage{bm}
\usepackage{subfigure}
\usepackage{natbib}
\usepackage{color}
\usepackage{multirow}

\newcommand{\beq}{\begin{equation}}
\newcommand{\eeq}{\end{equation}}
\newcommand{\bea}{\begin{eqnarray}}
\newcommand{\eea}{\end{eqnarray}}

\newcommand{\fb}{\;fb$^{-1}$}

\newcommand{\3}{4.9 cm}
\newcommand{\2}{7.0 cm}
\newcommand{\1}{12.5 cm}

\begin{document}

\preprint{CERN-PH-TH-2013-164} 

\title{\boldmath 
The role of interference in unraveling the $ZZ$-couplings  
of the newly discovered boson at the LHC}

\author{Mingshui Chen}
\author{Tongguang Cheng}
\author{James S. Gainer}
\email{Corresponding author: gainer@phys.ufl.edu}
\author{Andrey Korytov}
\author{Konstantin T. Matchev}  
\author{Predrag Milenovic}
\author{Guenakh Mitselmakher}
\affiliation{Physics Department, University of Florida, Gainesville, FL 32611, USA.}
\author{Myeonghun Park}
\email{Corresponding author: Myeonghun.Park@cern.ch}
\affiliation{CERN Physics Department, Theory Division, CH-1211 Geneva 23, Switzerland.}
\author{Aurelijus Rinkevicius}
\author{Matthew Snowball}
\affiliation{Physics Department, University of Florida, Gainesville, FL 32611, USA.}

\date{October 2, 2013} 

\begin{abstract}
We present a general procedure for measuring the tensor structure of the coupling
of the scalar Higgs-like boson recently discovered at the LHC to two
$Z$ bosons, including the effects of interference among different operators.
To motivate our concern with this interference, we explore the
parameter space of the couplings in the effective theory describing
these interactions and illustrate the effects of
interference on the differential dilepton mass distributions.
Kinematic discriminants for performing coupling measurements that
utilize the effects of interference are developed and described. We present
projections for the sensitivity of coupling measurements that use these
discriminants in future LHC operation in a variety of physics scenarios.
\end{abstract}

\maketitle


\clearpage
\pagebreak[4]

\section{Introduction}
\label{sec:intro}

As the new particle discovered by the ATLAS~\cite{Aad:2012tfa} and
CMS~\cite{Chatrchyan:2012ufa} collaborations appears to be
similar to the Standard Model (SM) Higgs boson~\cite{Freitas:2012kw,
  Corbett:2012ja, Chang:2013cia, Dumont:2013wma, Ellis:2013lra,
  Djouadi:2013qya}, it becomes very
important to measure its properties as precisely
as possible in order to find or constrain physics beyond the SM.
The recent ATLAS and CMS results strongly suggest that
the newly discovered boson has spin zero~\cite{
ATLAS-CONF-2013-013, CMS-PAS-HIG-13-002, 
ATLAS-CONF-2013-040, CMS-PAS-HIG-13-005, 
Aad:2013xqa},
which we take as the starting point in the studies presented in this report.
There is a large body of literature~\cite{Nelson:1986ki, Kniehl:1990yb, Soni:1993jc,
  Chang:1993jy, Barger:1993wt, Arens:1994wd, 
  Choi:2002jk, Buszello:2002uu, Schalla:2004ura, Godbole:2007cn, Kovalchuk:2008zz,
  Cao:2009ah, Gao:2010qx, DeRujula:2010ys, Englert:2010ud, Matsuzaki:2011ch, DeSanctis:2011yc,
  Gainer:2011xz, Englert:2012ct, Campbell:2012cz, Campbell:2012ct,
  Kniehl:2012rz, Moffat:2012pb, Coleppa:2012eh, Bolognesi:2012mm, Boughezal:2012tz,
  Stolarski:2012ps, Cea:2012ud, Kumar:2012ba, MEKD, Masso:2012eq,
  Chen:2012jy, Modak:2013sb,
  Kanemura:2013mc, Gainer:2013rxa, Heinemeyer:2013tqa,Caola:2013yja,
  Sun:2013yra,
  Anderson:2013fba} advocating the great potential
of $X \to ZZ \to 4\ell$ decays for disentangling the spin-parity
properties of resonances decaying to two Z bosons and for refining the
methodology for doing such measurements.   In this work
we explore the sensitivity of future LHC analyses to
interference between various operators in this channel. We follow
the framework of Ref.~\cite{Gainer:2013rxa}, which is briefly reviewed below.

\subsection{Review of Framework}

We consider a spin zero state $X$, which in general is a
linear combination of a $CP$-even state, $H$, and a  $CP$-odd state, $A$:
\beq
X \equiv H \cos\alpha + A \sin\alpha.
\label{Xdef}
\eeq
The couplings of the arbitrary spin zero boson, $X$, to two $Z$ bosons can be described by
the symmetry properties of the corresponding operators, which fall
into the following three categories: 
(i) $CP$-even terms which clearly violate gauge invariance,
(ii) $CP$-even terms which may preserve gauge invariance,
(iii) $CP$-odd terms.
For each category, the lowest dimensional operators in the
effective theory, in terms of some new physics scale $\Lambda$,
yield the Lagrangian
\beq
\mathcal{L} \supset
- \bigg(\frac{g_1 M_Z^2}{v}\bigg) H Z_\mu Z^\mu - 
\bigg(\frac{g_2}{2 \Lambda}\bigg) H F_{\mu\nu} F^{\mu\nu} -
\bigg(\frac{g_4}{2 \Lambda}\bigg) A F_{\mu\nu} \tilde{F}^{\mu\nu}, 
\label{lagrangian-g}
\eeq
where $\tilde{F}_{\mu\nu} = \frac{1}{2} \epsilon_{\mu\nu\rho\sigma}
F^{\rho\sigma}$ and the $g_i$ are dimensionless coupling constants.
Re-expressing the Lagrangian terms in Eq.~(\ref{lagrangian-g}) to
involve the mass eigenstate $X$, we obtain
\beq
\mathcal{L} \supset -X
\left[ \kappa_1 \frac{M_Z^2}{v} Z_\mu Z^\mu +
\frac{\kappa_2}{2 v} F_{\mu\nu} F^{\mu\nu} +
\frac{\kappa_3}{2 v} F_{\mu\nu}\tilde{F}^{\mu\nu}
\right],
\label{lagrangian}
\eeq
where
\beq
\kappa_1 = g_1 \cos{\alpha},~~~\kappa_2 = g_2 \cos{\alpha}
~(v/\Lambda),~~~\kappa_3 = g_4 \sin{\alpha}~(v/\Lambda).
\eeq
Each case where exactly one of the coefficients $\kappa_i$ is
non-vanishing corresponds to a specific pure
state:
(i) $\kappa_1 \neq 0$ corresponds to a SM-like Higgs (in particular $\kappa_1 = 1$ is the tree-level SM coupling);
(ii) $\kappa_2 \neq 0$ corresponds to the state which describes a SM singlet, usually denoted with $0^+_h$~\cite{Bolognesi:2012mm};
(iii) $\kappa_3 \neq 0$ corresponds to a pure pseudoscalar ($J^{CP} =0^-$).

The decay amplitude that one obtains from the Lagrangian in
Eq.~(\ref{lagrangian}) is
\begin{equation}
\mathcal{A} (X \to Z Z)  = 
 -\frac{2i}{v} \epsilon_{1}^{*\mu} \epsilon_{2}^{*\nu}
\left(
  (\kappa_1 M_{Z}^2 - \kappa_2 (p_1\cdot p_2)) g_{\mu \nu} 
+ \kappa_2 \,p_\mu p_\nu 
+ \kappa_3 \epsilon_{\mu\nu\alpha\beta}\,p_1^\alpha p_2^\beta
\right).
\label{amp} 
\end{equation}
Here $p_{1(2)}$ is the momentum of the intermediate $Z$ boson labelled
``$1$'' (``$2$''), while $p = p_1 + p_2$ is the momentum of the $X$ boson.
We note, following, e.g., Refs.~\cite{Gao:2010qx, Bolognesi:2012mm, DeRujula:2010ys}
(cf. especially Eq.~(11) in Ref.~\cite{Bolognesi:2012mm}) that the three
operators in Eq.~(\ref{lagrangian}) generate each of the three possible Lorentz
structures in the general amplitude for the decay of $X$ to two
bosons.  

\subsubsection{Comparison of Conventions}

Various conventions have
been used in writing Lagrangians and amplitudes for the study of the
$X \to ZZ$ interaction.  For the convenience of the reader,
Table~\ref{convention table} contains a dictionary of the
couplings used in Refs.~\cite{MEKD, Gao:2010qx,
  Bolognesi:2012mm, Djouadi:2013yb,Artoisenet:2013jma, Gainer:2013rxa}.

\begin{table}
\caption{\label{convention table} Comparison of notations for the effective $XZZ$ couplings.\footnote{
We note that an overall phase in the amplitude, which can be seen in
this table as an overall phase in the couplings, is irrelevant except in the
likely-negligible case of interference between, e.g., the $gg \to X \to
ZZ^\ast \to 4\ell$ signal, and the loop-induced $gg \to ZZ^\ast \to 4\ell$
background~\cite{Binoth:2008pr, Martin:2012xc, Kauer:2013cga,
  Dixon:2013haa}.}}
\begin{ruledtabular}
\begin{tabular}{cccc}
Ref.~\cite{Gainer:2013rxa} & $\kappa_1$ & $\kappa_2$ & $\kappa_3$ \\ \hline
Refs.~\cite{Gao:2010qx,Bolognesi:2012mm}  &
$(i/2) g^{(0)}_1$ & $-i g^{(0)}_2$ & $-i g^{(0)}_4$ \\
Ref.~\cite{MEKD} & $(g_{1z}/2)(v/M_Z^2)$ & $(g_{2z}/2)v$ & $(g_{4z}/2)v$ \\
Ref.~\cite{Djouadi:2013yb} & $gv/(2M_Z)$ & $g\lambda v/(2M_Z)$  & $-g\lambda' v/(2M_Z)$ \\
Ref.~\cite{Artoisenet:2013jma} & $-(g_{HZZ}k_{SM}v\cos\alpha)/(2M_Z^2)$ & $(k_{HZZ}v\cos\alpha)/(2\Lambda)$ & $(k_{AZZ}v\sin\alpha)/(2\Lambda)$ \\ 
\end{tabular}
\end{ruledtabular}
\end{table}

\subsubsection{Sensitivity to Loop-Induced Couplings}
\label{Subsec:loop}

The coefficients $\kappa_i$ in Eq.~(\ref{amp}) are real, since they originate from
the tree-level Lagrangian in Eq.~(\ref{lagrangian}).
By the optical theorem, the amplitude may obtain
contributions from loops with light particles (lighter than $M_X/2
\approx 63$ GeV) such that the expression for the amplitude including
loop effects is analogous to that in Eq.~(\ref{amp}), where the
effective couplings $\kappa^\prime_i$ are complex:
\begin{equation}
\mathcal{A} (X \to Z Z)  = 
 -\frac{2i}{v} \epsilon_{1}^{*\mu} \epsilon_{2}^{*\nu}
\left(
  (\kappa^\prime_1 M_{Z}^2 - \kappa^\prime_2 (p_1\cdot p_2)) g_{\mu \nu} 
+ \kappa^\prime_2 \,p_\mu p_\nu 
+ \kappa^\prime_3 \epsilon_{\mu\nu\alpha\beta}\,p_1^\alpha p_2^\beta
\right).
\label{amp prime}
\end{equation}

However, at least one of the $\kappa_i$ must not be predominantly loop-induced,
or else one runs into a contradiction with the experimental constrains.
For example, consider a generic loop with some invisible new particle whose coupling
to the $X (Z)$ boson is $g_{X(Z)}$.
Then, naively,
\beq
\delta \kappa^\prime_i =\frac{g_X g_Z^2}{16 \pi^2} \times \mathcal{O}(1).
\eeq
In this scenario the invisible width of the $X$ boson is
\beq
\Gamma_{X, \text{inv}} = \frac{g_X^2 M_X}{16 \pi} \times \mathcal{O}(1),
\eeq
hence taking $\Gamma_{X, \text{inv}} \lesssim
\Gamma_{X,\text{total}}^{exp} \lesssim 7$ GeV~\cite{CMS-PAS-HIG-13-016},
we obtain $g_X \lesssim 2$.  Since the gauge coupling to the $Z$,
$g_Z$, should be  $\lesssim 1$, we get
\beq
\delta \kappa^\prime_i \lesssim 1 \times 10^{-2}.
\eeq
This is about two orders of magnitude smaller than the magnitude of
couplings needed to give the SM rate~\cite{Gainer:2013rxa}. 
More stringent constraints on the $\delta \kappa^\prime_i$ (with some caveats) 
may be obtained from more stringent limits on the invisible width of the
Higgs~\cite{ATLAS-CONF-2013-011, ATLAS-CONF-2013-034,
  CMS-PAS-HIG-13-005,ATLAS:2013pma,Belanger:2013kya,Caola:2013yja} or the invisible
width of the $Z$~\cite{ALEPH:2005ab}.\footnote{We note that increasing the number of particles running in the
loop alleviates the constraints on the $\delta \kappa^\prime_i$.}
It is therefore well-motivated to treat the relevant couplings, $\kappa^\prime_i$,
(namely, the ones which are large enough to measure at present) as predominantly real.
  
\subsection{Experimental Situation}

The hypothesis of the new boson being a 100\% pure pseudoscalar, $0^-$,
has been excluded by CMS~\cite{Chatrchyan:2012jja, CMS-PAS-HIG-13-002} 
and ATLAS~\cite{ATLAS-CONF-2013-013, Aad:2013xqa}. 
The possibility of a 100\% pure $0^+_{\mathrm{h}}$ 
is also disfavored at 92\% C.L.~\cite{CMS-PAS-HIG-13-002}. 
Hence, in this study we assume a non-zero value of coupling, $\kappa_1$,
and address the question of the experimental sensitivity to the presence of
$\kappa_2$ and $\kappa_3$ terms in the $XZZ$ Lagrangian.

The current limit, set by CMS~\cite{CMS-PAS-HIG-13-002}, 
on the presence of a pseudoscalar contribution expressed 
in terms of a fractional cross section is
$f_{a3} =  \sigma_3 / ( \sigma_1 +  \sigma_3 ) < 0.58$.
Here the cross sections $\sigma_1$ and $\sigma_3$
are taken for the $4e$, $4\mu$, and $2e2\mu$ final states together\footnote
{
For given values of couplings $\kappa_i$ and $\kappa_j$,
the ratios of cross sections $\sigma_i / \sigma_j$~$(i \neq j)$
for same-fermion and different-fermion final states are different.
This is due to the interference effects associated with permutations of
identical fermions in the final state. 
Hence, one should specify which final states are used in the definition of $f_{a3}$. 
} and correspond to 100\% pure $0^+$ and $0^-$ states, respectively.
This result translates into a limit on the ratio of couplings $|\kappa_3 / \kappa_1| < 6.1$.
The corresponding CMS analysis was set up in such a way
that it was not sensitive to the interference between 
the $\kappa_3$- and $\kappa_1$-induced amplitudes.

\subsection{Objective}

In this paper, we show that by explicitly exploiting the
interference between amplitudes which involve the $0^+$, $0^+_{\mathrm{h}}$, and $0^-$ states
(corresponding to the $\kappa_1$, $\kappa_2$, and $\kappa_3$ terms in the
Lagrangian in Eq.~(\ref{lagrangian}), respectively)
one can boost the experimental sensitivities to the presence of an $0^-$ (and $0_h^+$) admixture. 
We show that the gains become particularly large at high integrated
luminosities, allowing one to probe smaller values of the
$\kappa_2$ and $\kappa_3$ couplings.
We also address the question of establishing the presence of the interference
and evaluating its sign, should decay amplitudes associated with spin zero 
higher dimensional operators be detected.
Recently, the importance of a proper treatment of interference
was discussed in the context of a somewhat different aspect 
of the $H \to ZZ \to 4\ell$ channel~\cite{MEKD}; there it was the interference 
associated with permutations of identical leptons in the $4e$ and $4\mu$
final states that was considered. This interference is always included
in the studies presented in this report.

\section{The physical importance of interference}

In general, interference effects can manifest themselves in two different ways:
either at the level of total cross sections (reflected in the production rate, as discussed in
Sec.~\ref{sec:rate} below), or at the level of differential distributions
(as discussed in Sec.~\ref{sec:distributions} below).

\subsection{The impact of interference effects on the production rate}
\label{sec:rate}

The overall rate for $X \to ZZ \to 4\ell$ events is proportional to the
partial width for $X\to ZZ$ \cite{Gainer:2013rxa}
\beq
\Gamma(X\to ZZ) = \Gamma_{SM} \sum_{i,j}\gamma_{ij}\kappa_i\kappa_j,
\label{GammaZZ}
\eeq
where the partial $H\to ZZ$ width predicted in the SM, $\Gamma_{SM}$, is factored out
in order to define constant dimensionless coefficients $\gamma_{ij}$ (with $\gamma_{ij}=\gamma_{ji}$) \footnote{The values quoted in Eq.~(\ref{gammas})
correspond to the $2e2\mu$ channel before cuts. For the $4e$ or $4\mu$
channels (or with cuts) the numerical values are similar but not identical~\cite{Gainer:2013rxa}.}
\beq
\gamma_{11}=1, \quad
\gamma_{22}=0.090, \quad
\gamma_{33}=0.038, \quad
\gamma_{12}=-0.250, \quad
\gamma_{13}=\gamma_{23}=0.
\label{gammas}
\eeq
The presence of interference is then implied by nonzero values of the ``off-diagonal''
coefficients $\gamma_{ij}$ with $i\ne j$. Eq.~(\ref{gammas}) shows that
the overall rate is affected by interference between $0^+$ and $0^+_{\mathrm{h}}$,
which is destructive (constructive) when $\kappa_1$ and $\kappa_2$
have the same (opposite) signs. Eq.~(\ref{gammas}) also implies
that at the level of total cross sections there is no interference between 
$0^+$ and $0^-$ or between $0^+_{\mathrm{h}}$ and $0^-$.

The magnitude of interference depends on the values of the 
couplings $\kappa_1$ and $\kappa_2$. Obviously, for a pure $0^+$
state ($\kappa_1\ne 0$, $\kappa_2=0$) and for a pure $0^+_{\mathrm{h}}$
state ($\kappa_2\ne 0$, $\kappa_1=0$) the interference is absent.
Given the values in Eq.~(\ref{gammas}), one could expect 
the interference effect to be maximal for
\beq
\frac{\kappa_2}{\kappa_1} = 
\frac{1}{2} \tan^{-1}\bigg( \frac{2 \gamma_{12}}{\gamma_{1}-\gamma_{22}}\bigg)
\simeq 3.89.
\label{maxinterference}
\eeq

In practice, the signal rate for $X \to ZZ \to 4\ell$ production is measured from data,
thus imposing one constraint through Eq.~(\ref{GammaZZ}) on the $\{\kappa_1,\kappa_2,\kappa_3\}$ 
parameter space \cite{Gainer:2013rxa} (provided the production rate
for the $X$ is fixed). The constraint may be solved explicitly by a suitable 
change of variables, reducing the relevant $\{\kappa_i\}$ parameter space to a two-dimensional surface
which can be taken to be effectively the surface of a sphere \cite{Gainer:2013rxa}.
For this reason, we shall not discuss the overall rate further. Instead, we will assume in our
analyses that the rate measurement has already been performed and the
couplings $\kappa_i$ have been chosen so that they satisfy the
constraint of Eq.~(\ref{GammaZZ}).

\subsection{The impact of interference effects on differential distributions}
\label{sec:distributions}

Even if the overall rate is kept fixed, the interference effects are still present 
at the level of differential distributions
(the size of this effect will be quantified in Sec.~\ref{sec:results} below).
In general, the kinematics of $X \to ZZ \to 4\ell$ events is described in the $X$ rest frame by 
7 independent degrees of freedom, and interference will impact the differential 
distribution in this 7-dimensional signature space. For simplicity, 
in this subsection we will focus only on the $M_{Z_1}$ and $M_{Z_2}$ 
invariant mass distributions and use them to illustrate the effects of interference.\footnote{Note
the webpage {\tt http://yichen.me/project/GoldenChannel/} created by
the authors of Ref.~\cite{Chen:2012jy}, may be used to make plots of
interesting differential distributions for different values of the
couplings $\kappa_i$.}
In order to provide an intuitive understanding of some of the results to follow in 
Sec.~\ref{sec:results}, we shall derive analytical formulas for the $M_{Z_1}$ 
and $M_{Z_2}$ distributions, which explicitly demonstrate the interference effects.

The doubly differential decay width with respect to $M_{Z_1}$ and 
$M_{Z_2}$ can be written as
\begin{equation}
\frac{d^2\, \Gamma}{d M_{Z_1} d M_{Z_2}} = \frac{1}{v} \sum_{i,j} \kappa_i \kappa_j
F_{ij} (M_{Z_1}, M_{Z_2}; M_X),
\label{dGdMZ1dMZ2}
\end{equation}
where the dimensionless\footnote{Since $\kappa_i$ are already dimensionless,
in the right-hand side of Eq.~(\ref{dGdMZ1dMZ2}) we factor out
$1/v$ to make $F_{ij}$ dimensionless as well.} functions $F_{ij}$ are symmetric with respect to their indices:
$F_{ij} = F_{ji}$.
In the absence of any selection criteria, the functions $F_{ij}$ are 
\begin{eqnarray}\label{eq:Fij}
F_{11} (M_{Z_1}, M_{Z_2}) &=& \frac{M_Z^4}{M_{Z_1}^2 M_{Z_2}^2}(x+3)\, \xi(M_{Z_1}, M_{Z_2}; M_X), \label{F11}\\ [2mm]
F_{12} (M_{Z_1}, M_{Z_2}) &=& \frac{M_Z^2}{M_{Z_1} M_{Z_2}}\, 3\sqrt{x+1}\, \xi(M_{Z_1}, M_{Z_2}; M_X), \label{F12}\\ [2mm]
F_{22} (M_{Z_1}, M_{Z_2}) &=& (2x+3)\, \xi(M_{Z_1}, M_{Z_2}; M_X),\label{F22}\\[2mm]
F_{13} (M_{Z_1}, M_{Z_2}) &=& 0,\label{F13}\\[2mm]
F_{23} (M_{Z_1}, M_{Z_2}) &=& 0,\label{F23}\\[2mm]
F_{33} (M_{Z_1}, M_{Z_2}) &=& 2x\, \xi(M_{Z_1}, M_{Z_2}; M_X), \label{F33}
\end{eqnarray}
where the dimensionless common factor $\xi$ is given by 
\beq
\xi(M_{Z_1}, M_{Z_2}; M_X)\equiv \left(\frac{g_2^2(g_a^2+g_v^2) }{2\cos\theta_W^2}\right)^2
\frac{M_{Z_1}^6 M_{Z_2}^6 \sqrt{x} }{9(2\pi)^5\, v\, M_X^3}
\frac{1 }{P_1 P_2}.
\eeq
Here $g_2$ is the $SU(2)_W$ gauge coupling constant, 
$g_v = -\frac{1}{2}+2 \sin\theta_W^2$, $g_a = -\frac{1}{2}$,
$\theta_W$ is the Weinberg angle,
\beq
x \equiv  \left(\frac{M_X^2-M_{Z_1}^2-M_{Z_2}^2}{2 M_{Z_1} M_{Z_2}}\right)^2-1
\eeq
is a dimensionless parameter introduced in Ref.~\cite{Bolognesi:2012mm}, and 
\beq
P_i \equiv (M_{Z_i}^2-M_Z^2)^2 + \Gamma_Z^2 \,M_Z^2
\label{props}
\eeq
are the $Z$ propagator functions which depend on the mass, $M_Z$, and
width, $\Gamma_Z$, of the $Z$-boson.

The doubly differential distribution in Eq.~(\ref{dGdMZ1dMZ2})
is an interesting object to study experimentally and 
CMS and ATLAS have published plots of the Higgs candidate events in the
$(M_{Z_1},M_{Z_2})$ plane. Events are expected to be clustered around $M_{Z_1}=M_Z$,
while the $M_{Z_2}$ dependence is non-trivial and contains interesting information~\cite{MEKD}.
Therefore we integrate the expression in
Eq.~(\ref{dGdMZ1dMZ2}) over $M_{Z_1}$ and consider instead the
corresponding one dimensional distribution 
\bea
\frac{d\, \Gamma}{d M_{Z_2}} \equiv \int d M_{Z_1} \left( \frac{d^2\, \Gamma}{d M_{Z_1} d M_{Z_2}} \right)
\equiv \sum_{i,j} \kappa_i \kappa_j f_{ij} (M_{Z_2}; M_X),
\label{dNdMZ21dim}
\eea
with newly defined dimensionless functions 
\beq
f_{ij} (M_{Z_2}; M_X) \equiv \frac{1}{v}\int d M_{Z_1}\, F_{ij} (M_{Z_1}, M_{Z_2}; M_X)
\label{smallfij}
\eeq
in place of Eqs.~(\ref{F11}-\ref{F33}). Comparison of
Eq.~(\ref{GammaZZ}) and Eq.~(\ref{dGdMZ1dMZ2}) shows that
the normalization of the functions $F_{ij}$ and $f_{ij}$ 
is given by the values of the coefficients $\gamma_{ij}$ in Eq.~(\ref{gammas})~\cite{Gainer:2013rxa} 
\bea
  \gamma_{ij} &=& \frac{1}{v\, \Gamma_{SM}} \int d M_{Z_1} \int d M_{Z_2} \, F_{ij} (M_{Z_1}, M_{Z_2}; M_X) \\ [2mm]
                      &=& \frac{1}{\Gamma_{SM}} \int d M_{Z_2}\,  f_{ij} (M_{Z_2}; M_X).
\label{fijnorm}                      
\eea

\begin{figure}[t]
\begin{center}
\includegraphics[width=7cm]{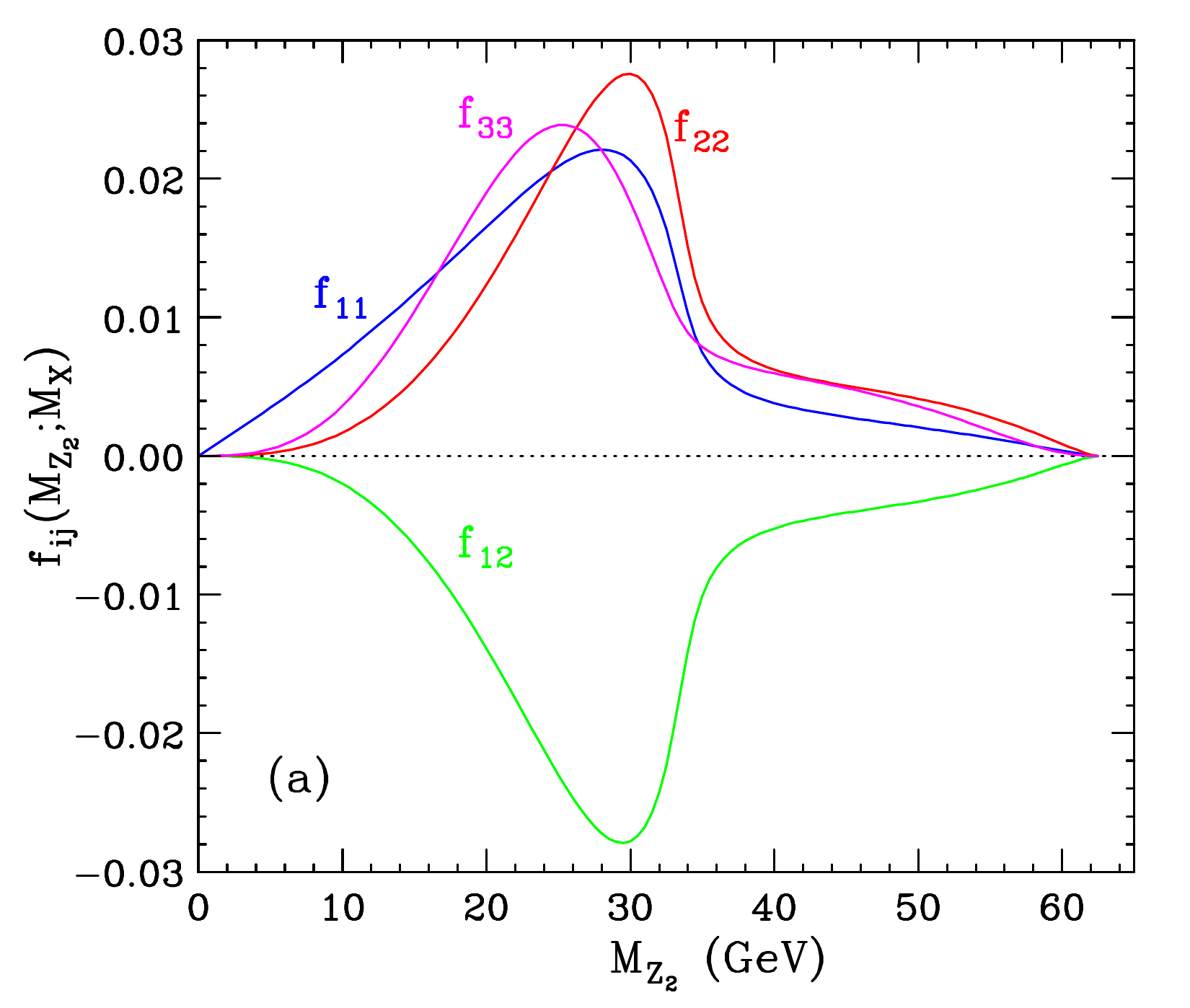} 
\includegraphics[width=7.4cm]{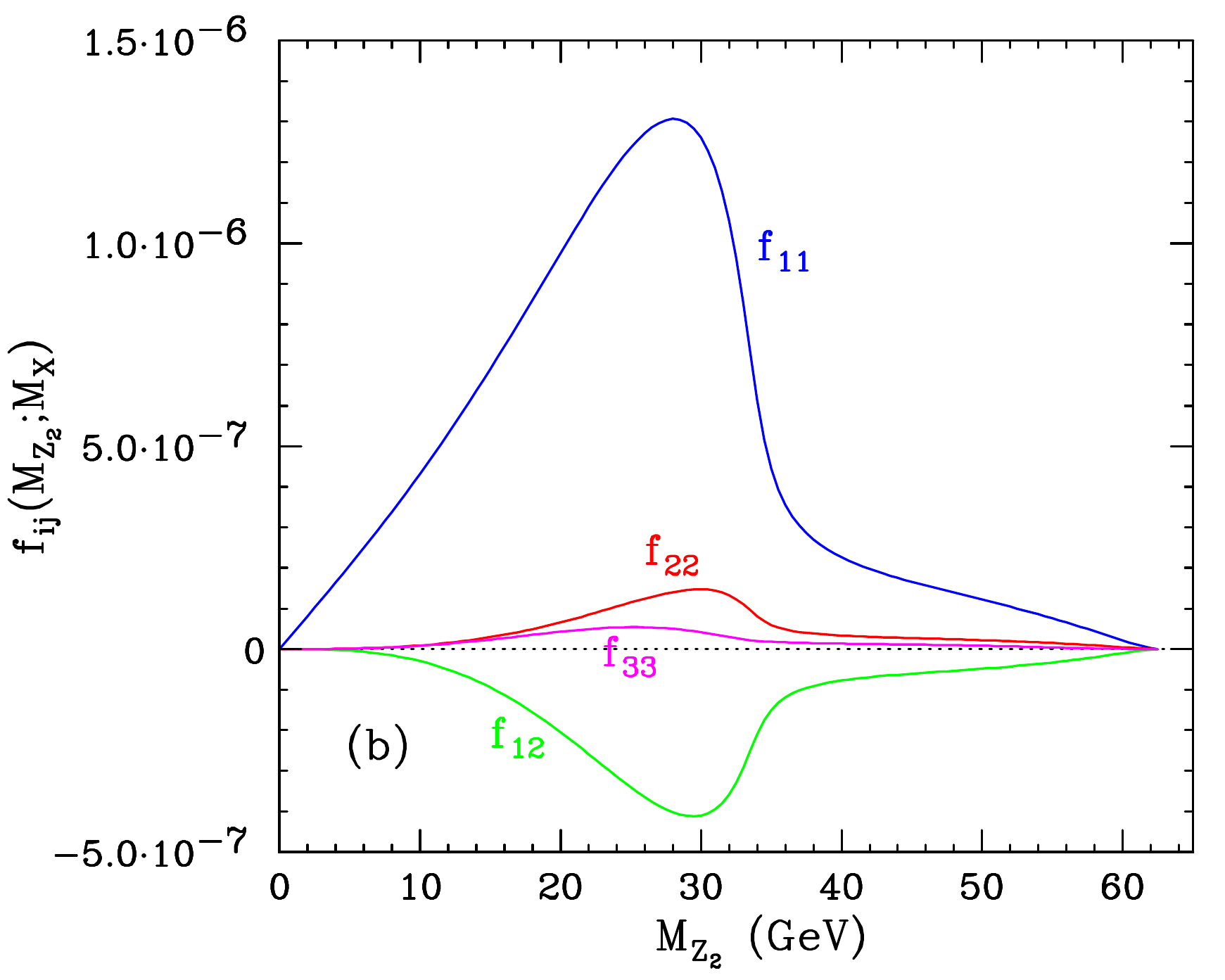} 
\end{center}
\caption{The four non-vanishing functions $f_{ij}$ defined in Eq.~(\ref{smallfij}) 
as a function of $M_{Z_2}$, with $M_X=125$ GeV and (a) unit normalization
or (b) properly normalized as in Eq.~(\ref{fijnorm}).
\label{fig:Fij}}
\end{figure} 

Fig.~\ref{fig:Fij} shows the four non-vanishing functions
$f_{11}$ (blue), $f_{12}$ (green), $f_{22}$ (red) and $f_{33}$ (magenta)
as a function of $M_{Z_2}$ for the nominal value of $M_X=125$ GeV. 
The two functions $f_{13}$ and $f_{23}$ vanish due to the $CP$ properties of the operators
considered in the Lagrangian in Eq.~(\ref{lagrangian}).
All functions in panel (a) are normalized to unity, which makes it easier to
study the differences in their shapes. In panel (b) the functions are 
properly normalized in accordance with Eq.~(\ref{fijnorm}).

Fig.~\ref{fig:Fij}(a) shows that all four functions exhibit similar dependance on $M_{Z_2}$.
At first, they all monotonically increase from $0$ at $M_{Z_2}=0$,
reaching a peak somewhere in the neighborhood of $M_{Z_2}\sim 25-30$ GeV,
followed by a sudden drop at around $M_{Z_2}\sim 34$ GeV, and a long tail
until $M_{Z_2} =62.5$ GeV.
This behavior can be understood purely in terms of kinematics.
The majority of the events contain an on-shell $Z$-boson with $M_{Z_1}\approx M_Z$,
which leaves only up to $M_X-M_Z\sim 34$ GeV available to $M_{Z_2}$, which 
explains the kinematic endpoint at $M_{Z_2}\sim 34$ GeV. The tail results from 
events where both $Z$-bosons are off-shell, and extends to half the $X$ mass,
$M_X/2=62.5$ GeV. Finally, the distributions peak relatively close to the 
$M_{Z_2}\sim 34$ GeV endpoint, since the propagator functions in Eq.~(\ref{props})
prefer $M_{Z_2}$ to be as close as possible to the mass $M_Z$ of the 
$Z$-boson\footnote{Contrast this to the case of the SM background, where there is a contribution 
from a virtual photon which dominates and causes the $M_{Z_2}$ distribution to
peak at much lower values \cite{Bolognesi:2012mm,MEKD}.}.

Fig.~\ref{fig:Fij}(b) compares the relative size of the different $f_{ij}$ functions.
We see that the overall magnitude is largest for $f_{11}$ and smallest for $f_{33}$.
Note that the interference contribution from $f_{12}$ has the second largest magnitude 
and an opposite sign compared to the other three functions shown in the plot ---
these two facts will be important in the discussion to follow.

The observable $M_{Z_2}$ distribution is obtained by a suitable superposition of the
individual contributions seen in Fig.~\ref{fig:Fij}(b), properly weighted by products of 
$\kappa_i$ couplings as specified in Eq.~(\ref{dNdMZ21dim}). Fig.~\ref{fig:Fij} allows us
to understand the resulting $M_{Z_2}$ shapes. First, we concentrate on the location 
of the peak of the total $M_{Z_2}$ distribution, which has been suggested as an
easily measurable global observable characterizing any invariant mass distribution
\cite{Cho:2012er}. 
\begin{figure}[t]
\includegraphics[width=\1]{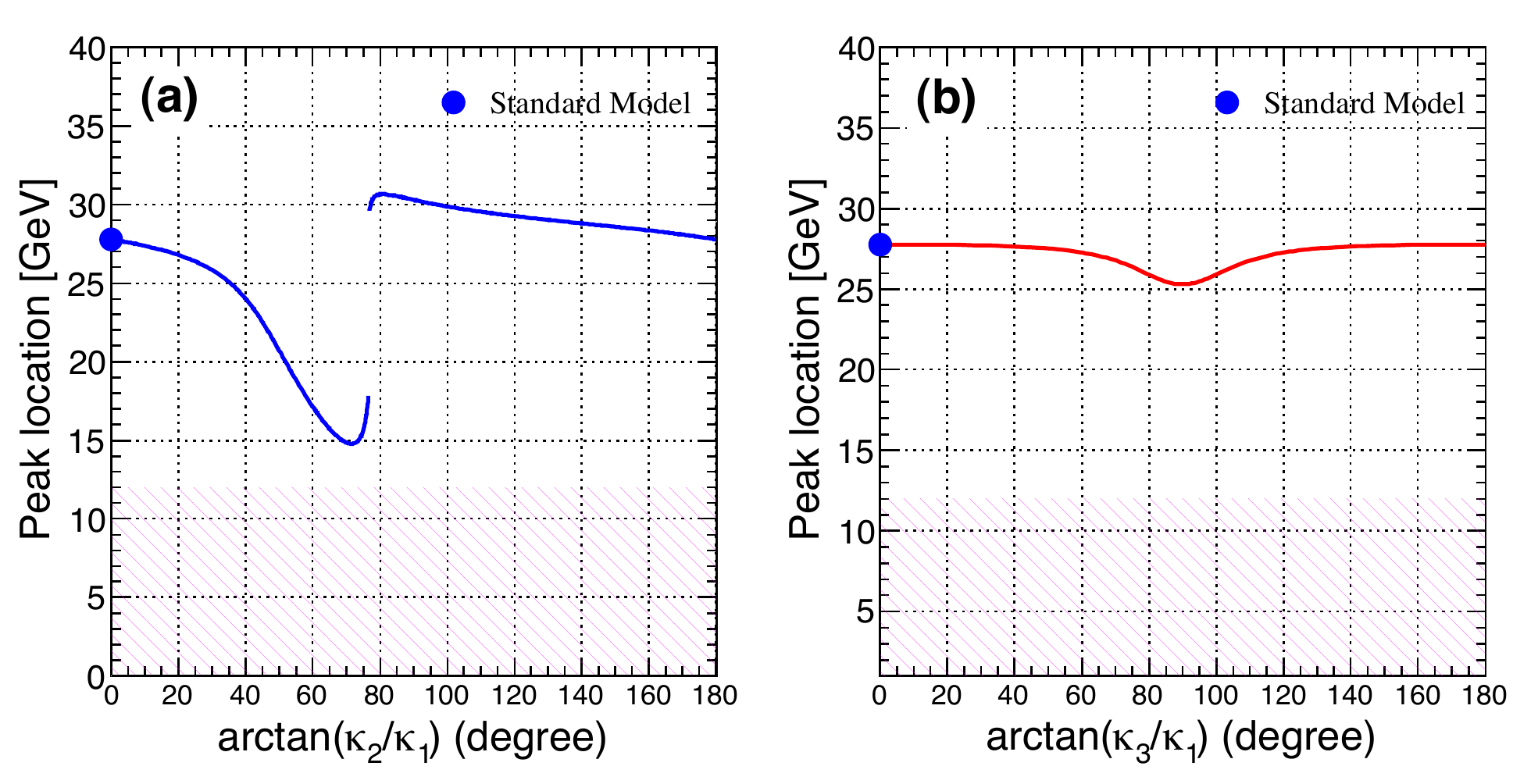} 
\caption{The location of the peak in the $M_{Z_2}$ distribution
as a function of 
(a) the ratio $\kappa_2/\kappa_1$, with $\kappa_3=0$ and 
(b) the ratio $\kappa_3/\kappa_1$, with $\kappa_2=0$. 
The shaded region denotes the lower cut on $M_{Z_2}$ used in our analysis.
The blue circle corresponds to the case of the tree-level SM ($\kappa_1=1$, $\kappa_2= 0$, $\kappa_3= 0$).
\label{fig:peak}}
\end{figure} 
The peak location is plotted in Fig.~\ref{fig:peak} for two scenarios:
(a) $\kappa_3=0$ and varying the ratio $\kappa_2/\kappa_1$, keeping the total 
$X \to ZZ \to 4\ell$ partial width fixed to $\Gamma_{SM}$; and
(b) $\kappa_2=0$ and similarly varying the ratio $\kappa_3/\kappa_1$.

Let us first focus on the interplay between the $\kappa_1$ and $\kappa_3$
terms in the Lagrangian (\ref{lagrangian}). In this case, the behavior of the peak
shown in Fig.~\ref{fig:peak}(b) is relatively simple, due to
the absence of an interference contribution ($f_{13}=0$).
The case of the SM (denoted by the blue circle) corresponds to $\kappa_1=1$ and $\kappa_3=0$,
in which case the $M_{Z_2}$ distribution is made up entirely of 
the $f_{11}$ contribution, which peaks around 28 GeV.
As the value of $\kappa_3$ is gradually increased, one introduces a larger fraction of
the $f_{33}$ component from Fig.~\ref{fig:Fij}, which peaks at a lower value of
$M_{Z_2}$, around 25 GeV. 
As a result, the peak location in Fig.~\ref{fig:peak}(b) 
is initially a decreasing function of the ratio $\kappa_3/\kappa_1$.
Eventually, we reach the case of a pure $0^-$ state with $\kappa_2\ne 0$ and $\kappa_1=0$,
when the $M_{Z_2}$ distribution is composed entirely of the $f_{33}$ component and the
$M_{Z_2}$ peak is located at $M_{Z_2}\sim 25$ GeV.
The right half of Fig.~\ref{fig:peak}(b), where the $\kappa_1$ and $\kappa_3$ couplings
are taken with a relative minus sign, is a mirror image of the left
and can be understood in the same way.

Notice that the $f_{11}$ and $f_{33}$ contributions always enter with positive weights, 
$\kappa_1^2$ and $\kappa_3^2$, respectively. Thus the shape of the combined
$M_{Z_2}$ distribution is a weighted average between the $f_{11}$ and $f_{33}$
shapes seen in Fig.~\ref{fig:Fij}(a), which are already very similar.
As a result, the peak location stays relatively constant
over the whole range of the couplings ratio $\kappa_3/\kappa_1$.

In contrast, when we consider the interplay between $\kappa_1$ and 
$\kappa_2$, the situation changes completely, as demonstrated by Fig.~\ref{fig:peak}(a).
Now the $M_{Z_2}$ distribution is built up from three components:
$f_{11}$, which peaks near 28 GeV, $f_{22}$, which peaks around 30 GeV, 
and $f_{12}$, whose magnitude peaks near 29 GeV.   
Given that the peaks of all these three components are very close,
one might expect that the peak of the total $M_{Z_2}$ distribution 
would also fall in the vicinity of $28-30$ GeV. However, Fig.~\ref{fig:peak}(a)
reveals that this naive expectation is false and that in the range where 
the couplings $\kappa_1$ and $\kappa_2$ have the same sign, 
the peak location can vary
from as low as $15$ GeV 
to as high as $31$ GeV. The reason for this wild behavior can be traced 
to the fact that the interference term, $f_{12}$, is significant and 
{\em opposite in sign} from $f_{11}$ and $f_{22}$, so that when
the couplings $\kappa_1$ and $\kappa_2$ have equal signs, it
destructively interferes with the sum of the $f_{11}$ and $f_{22}$ terms.
Even more surprisingly, as the value of $\kappa_2$ is increased relative to
$\kappa_1$, at a certain point the $M_{Z_2}$ distribution undergoes a type 
of ``first order phase transition", where the location of the peak ``jumps" 
suddenly and discontinuously from around $18$ GeV to near $30$ GeV,
signaling the presence of at least two local maxima in the $M_{Z_2}$
distribution.

\begin{figure}[t]
\begin{center}
\includegraphics[width=\3]{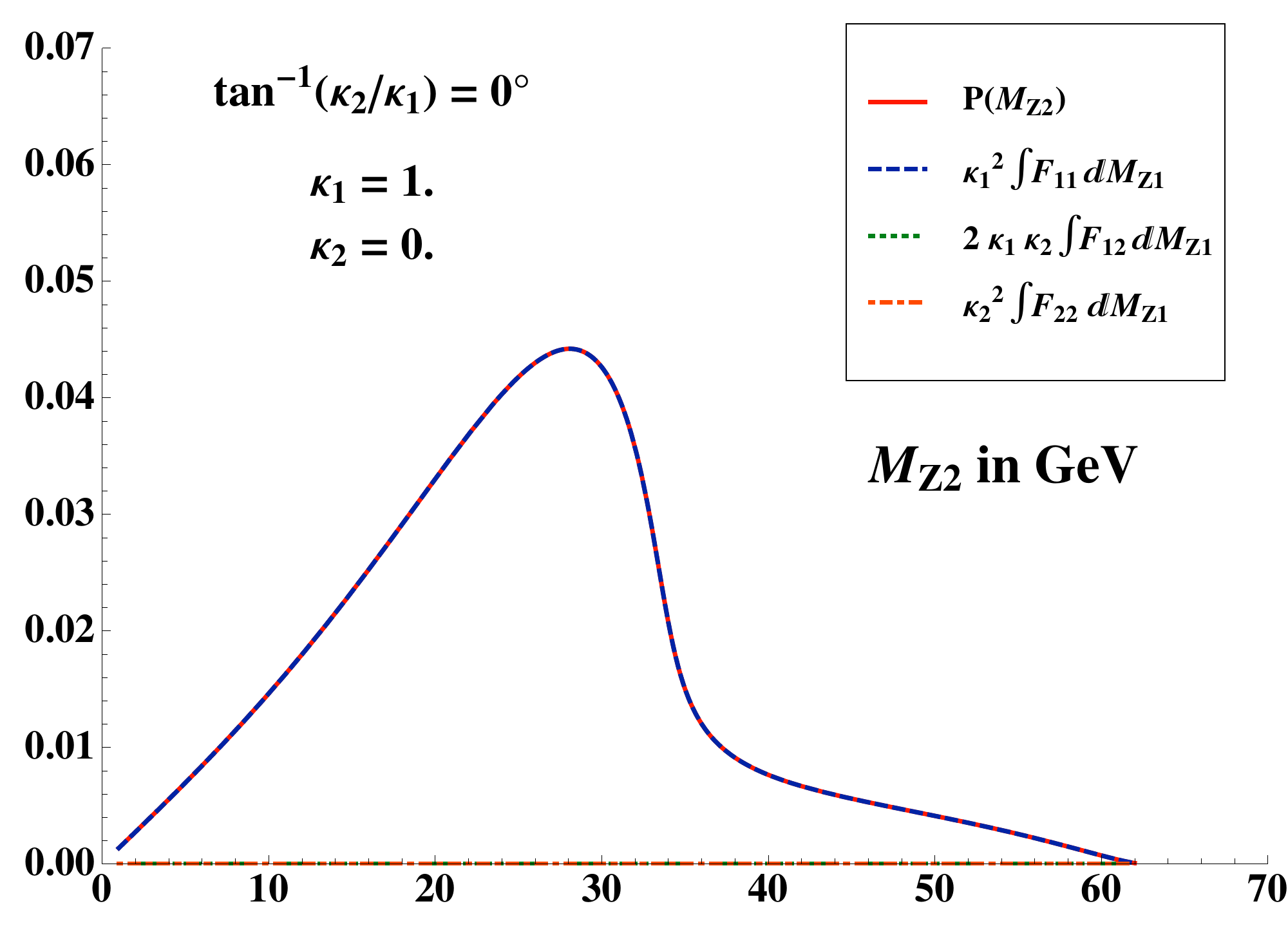} ~~
\includegraphics[width=\3]{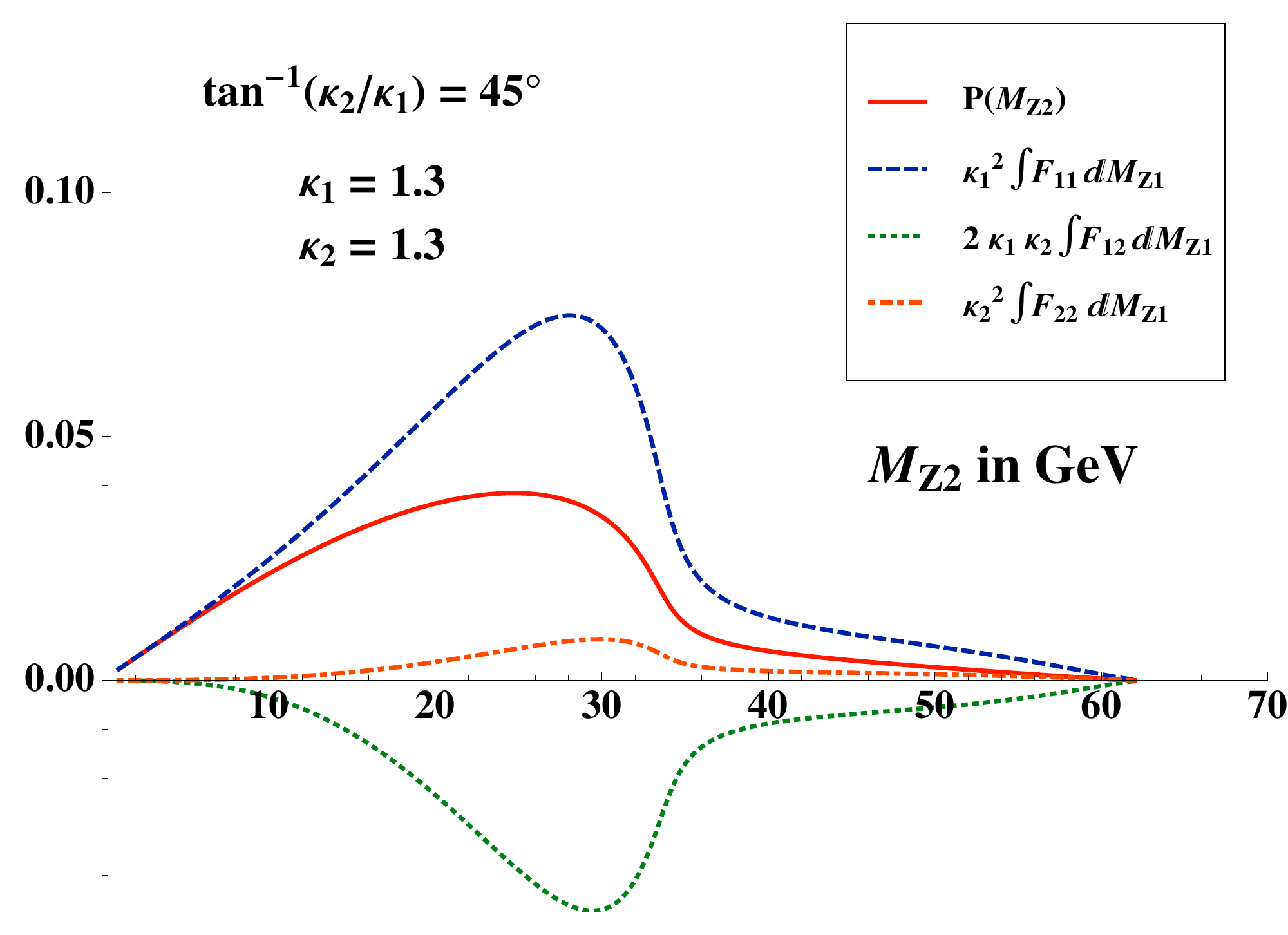} ~~
\includegraphics[width=\3]{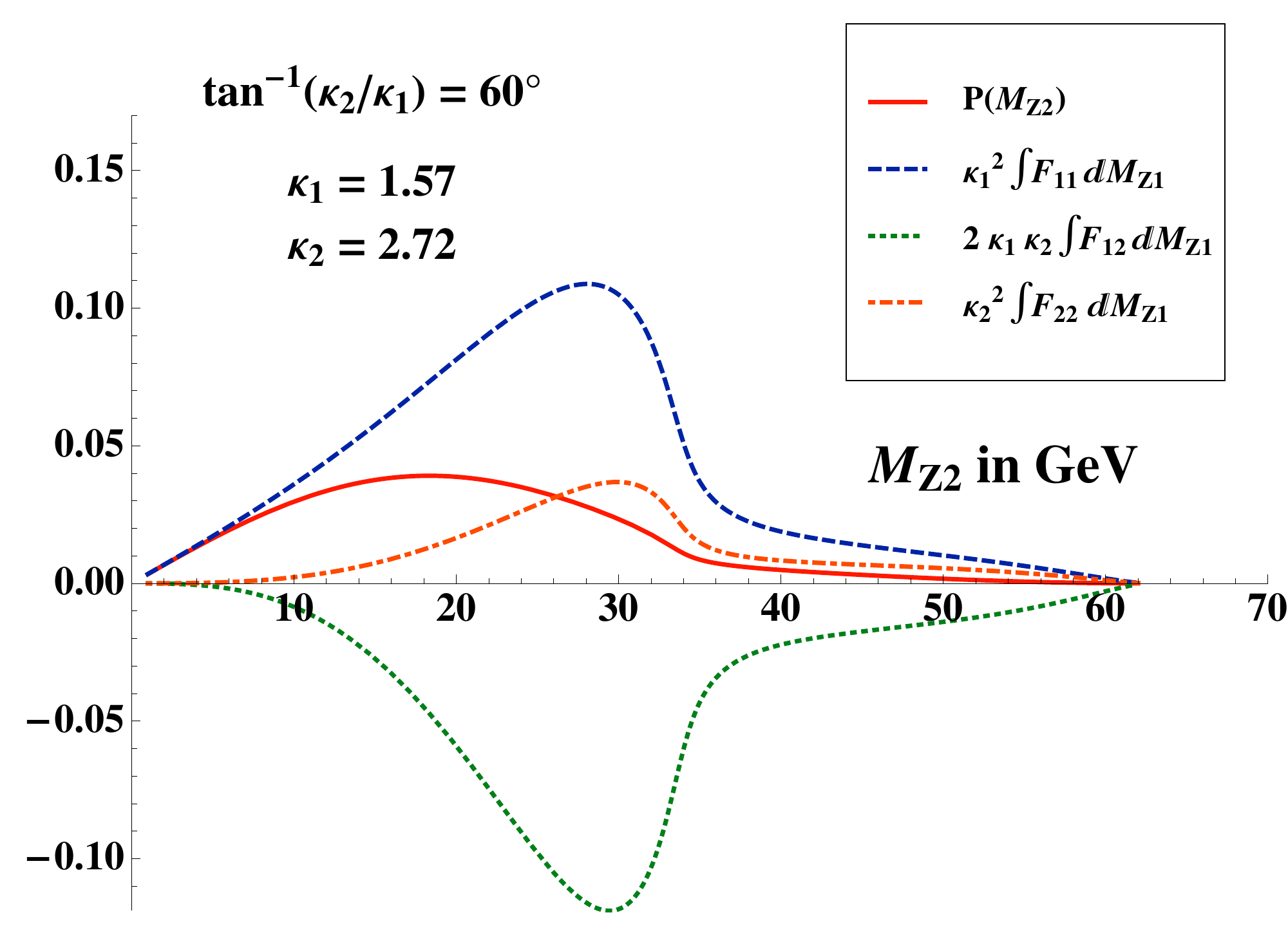} \\
\includegraphics[width=\3]{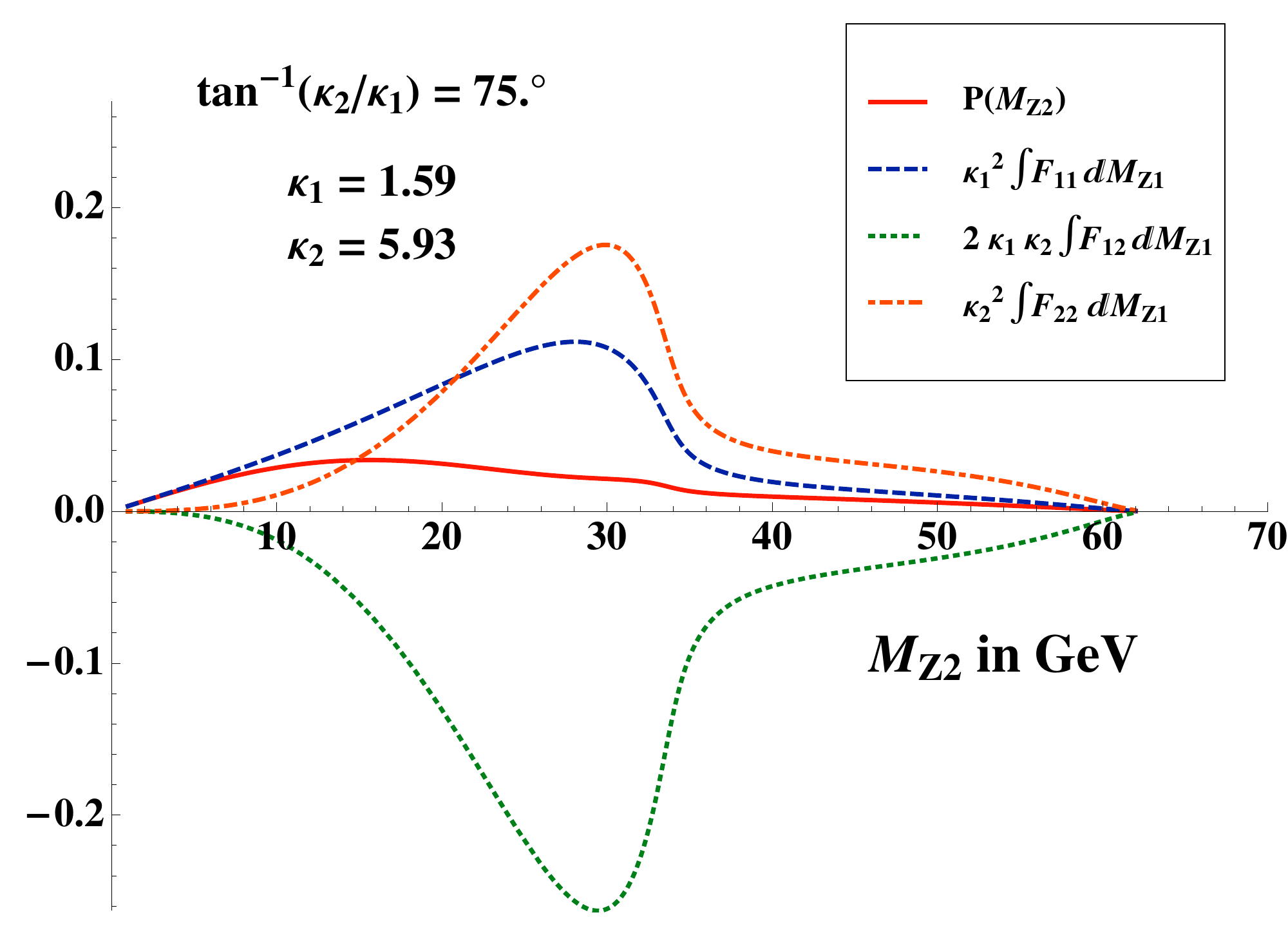} ~~
\includegraphics[width=\3]{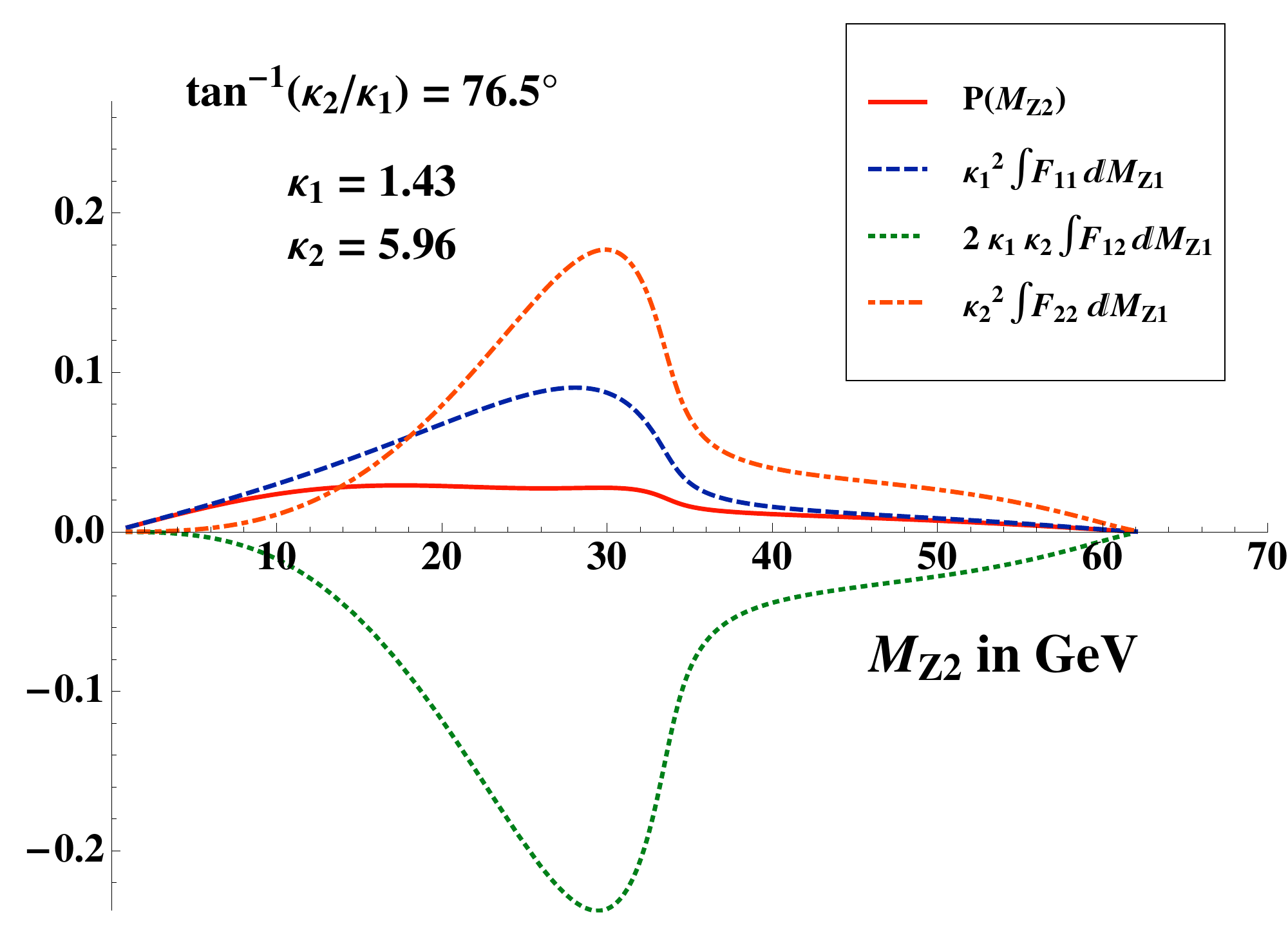} ~~
\includegraphics[width=\3]{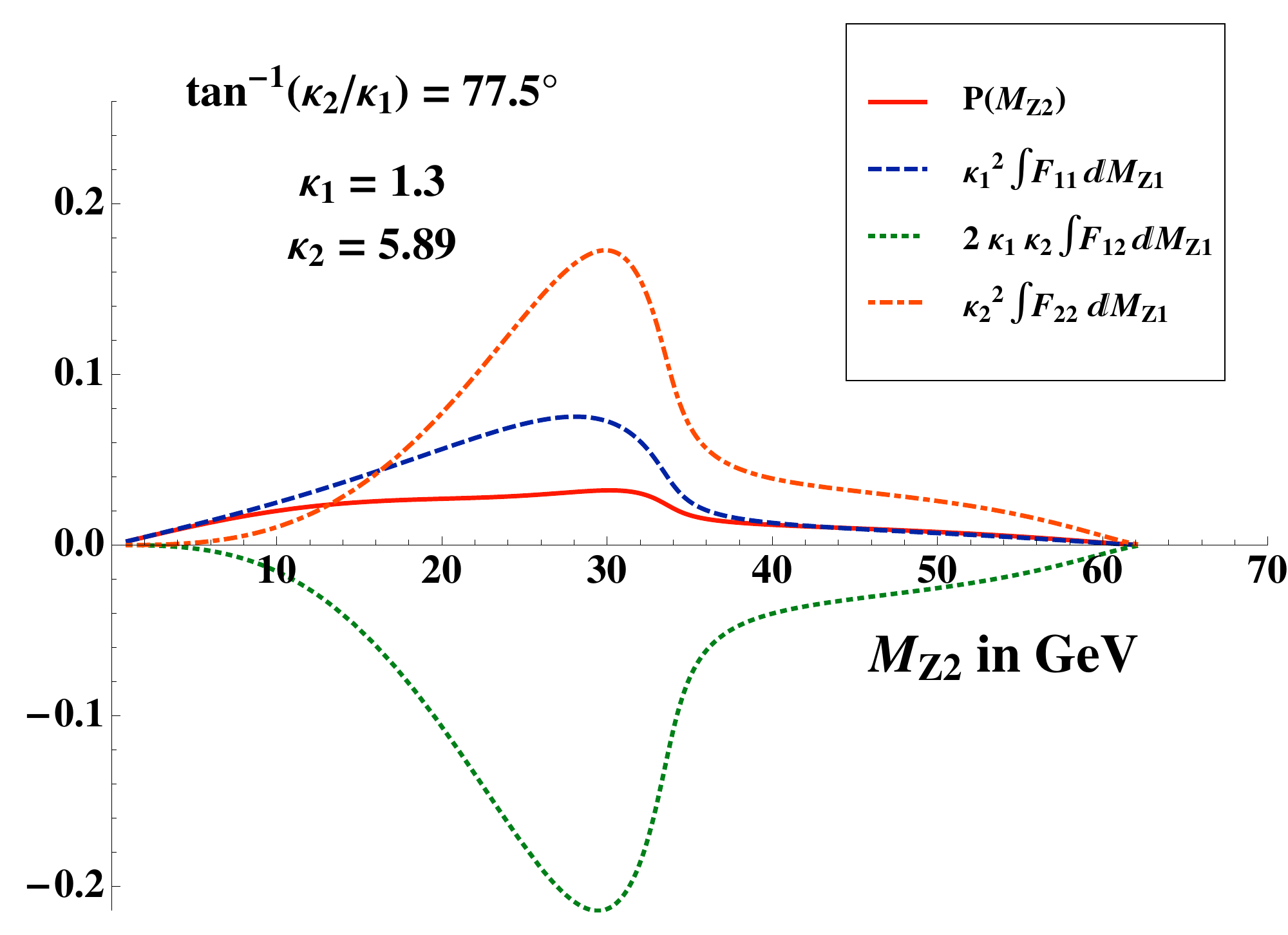} \\
\includegraphics[width=\3]{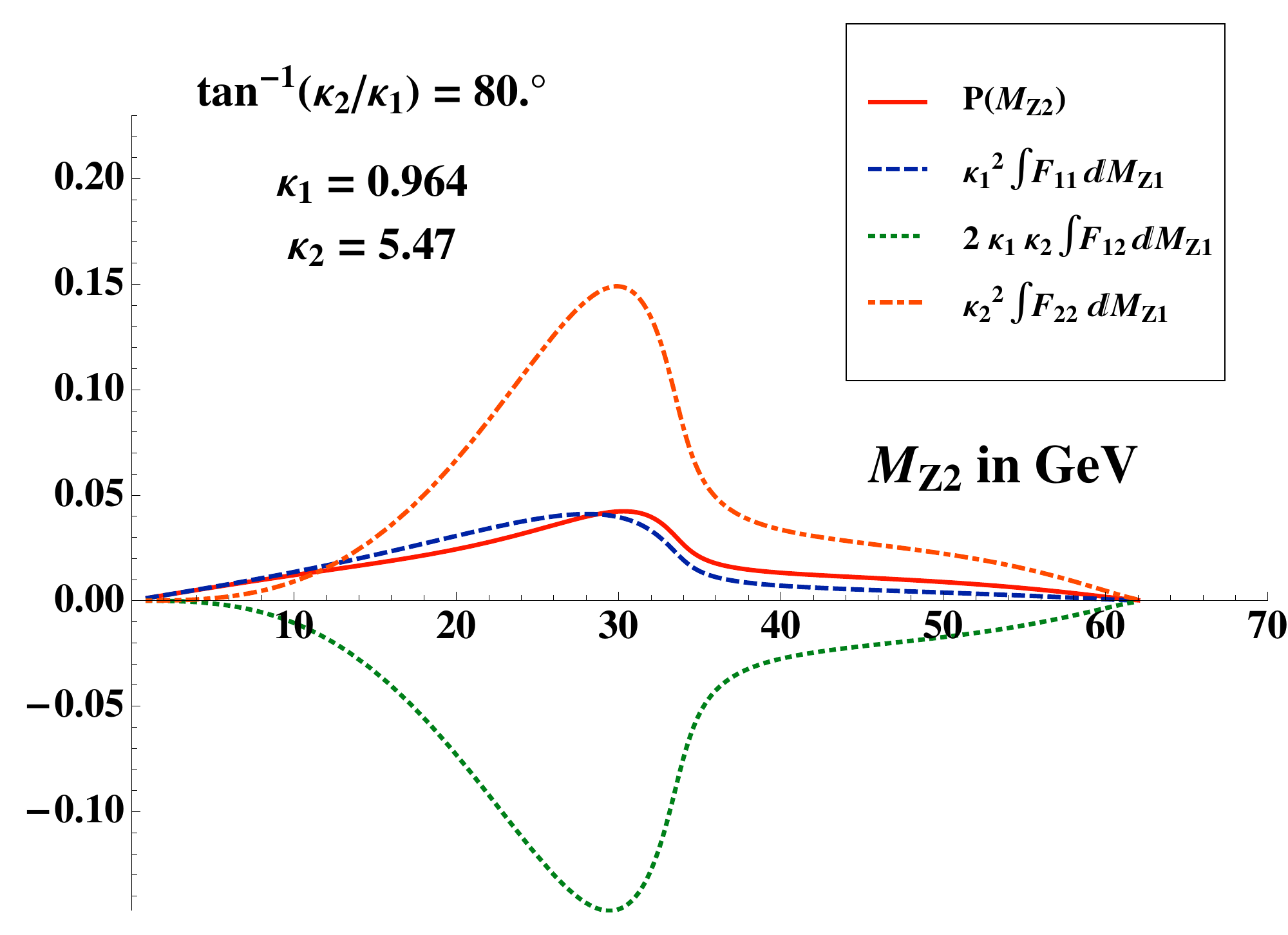} ~~
\includegraphics[width=\3]{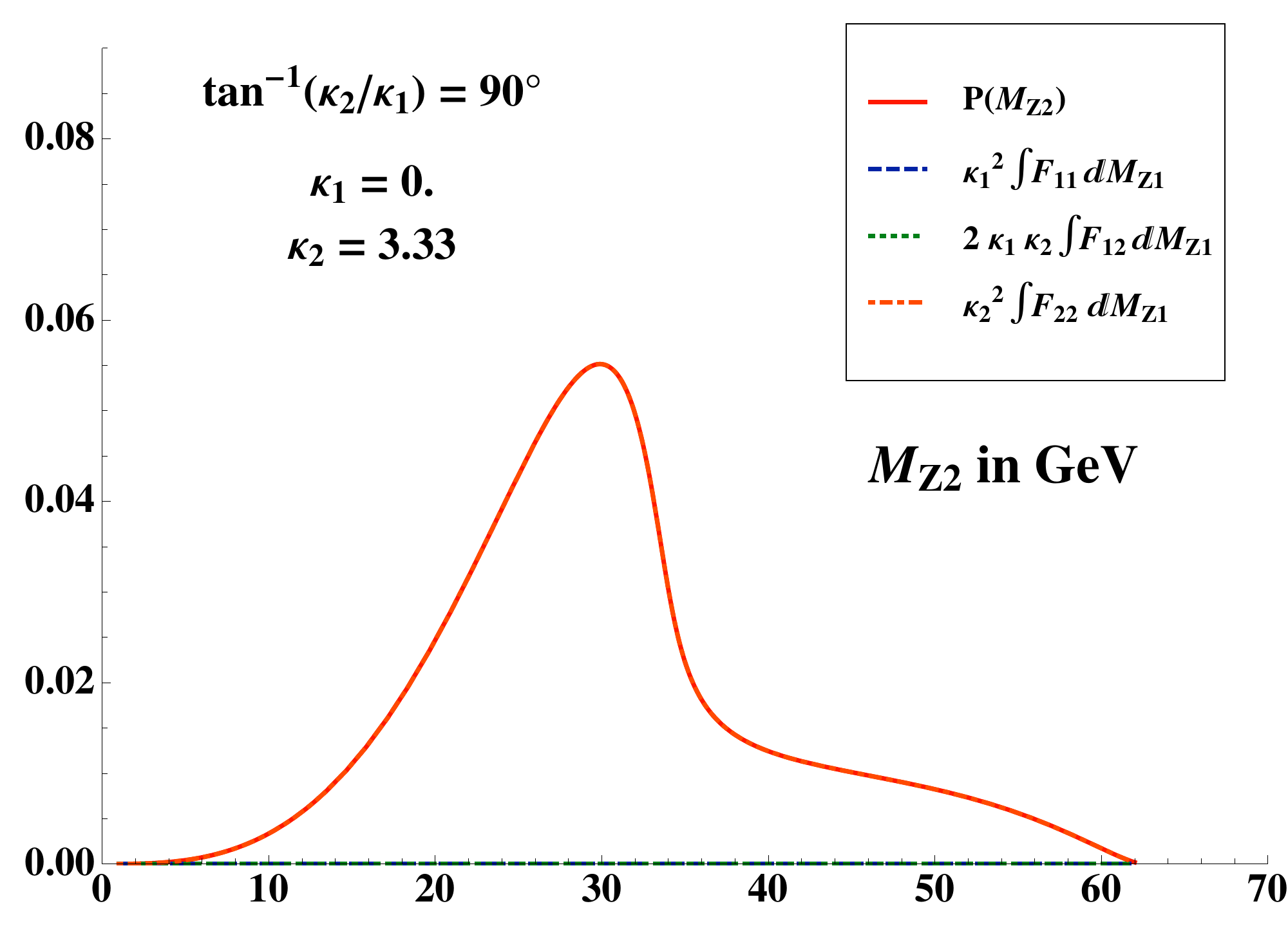} ~~
\includegraphics[width=\3]{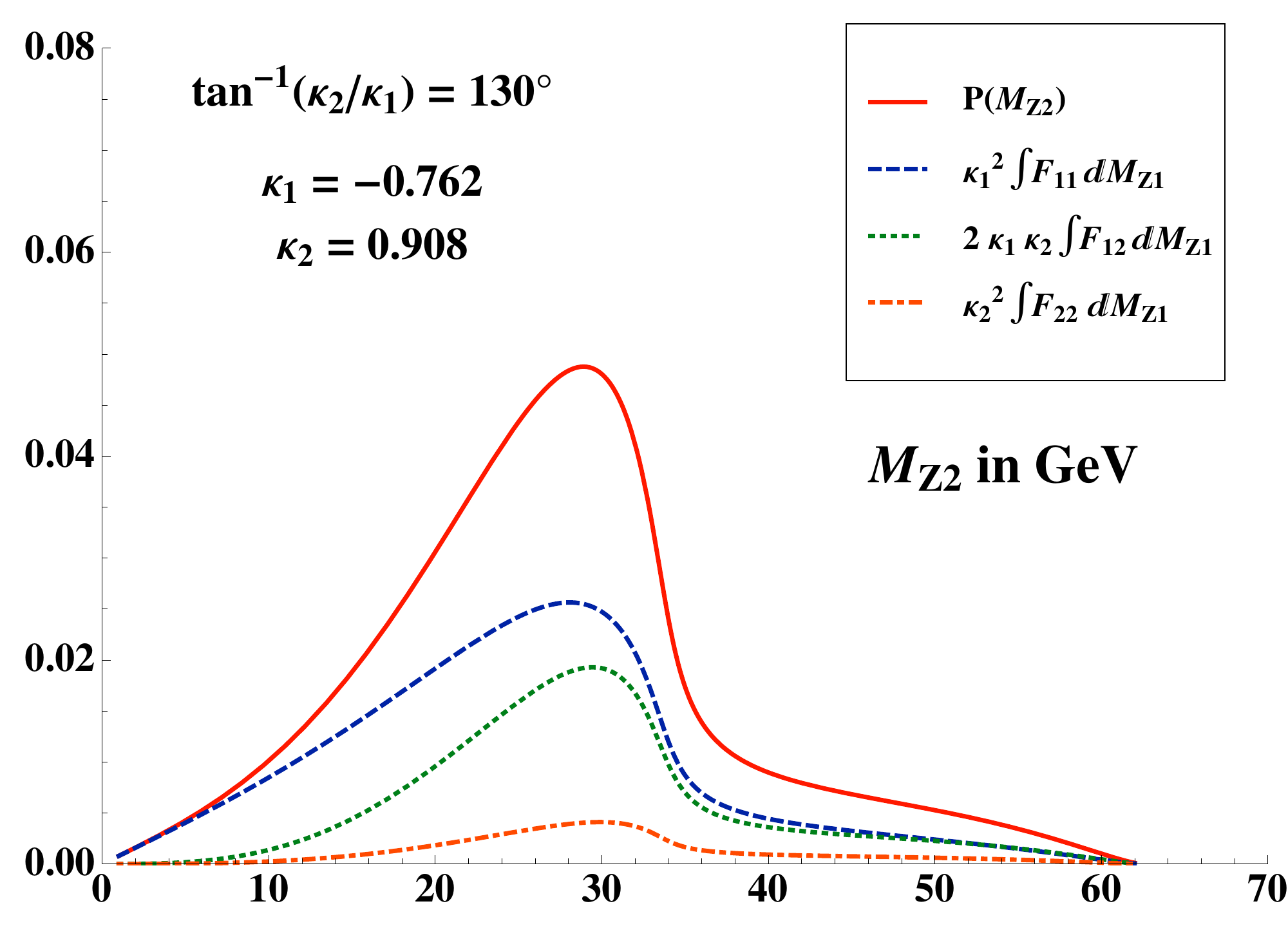} 
\end{center}
\caption{$M_{Z2}$ distributions for $\kappa_3=0$ and different choices of $\kappa_1$ and $\kappa_2$. 
The net total (shown in solid red) is comprised of three contributions: from $f_{11}$ (dashed blue),
from $f_{22}$ (dot-dashed orange), and from the interference term
$f_{12}$ (dotted green).  More plots like these are available in movie
form at \texttt{http://www.phys.ufl.edu/\textasciitilde gainer/k1k2-movie.mov}.
}
\label{fig:MZ2movie}
\end{figure} 

The peculiarities exhibited in Fig.~\ref{fig:peak}(a) prompt further detailed investigations. 
In Fig.~\ref{fig:MZ2movie} we plot the $M_{Z_2}$ distribution (shown with a red solid line)
for a series of interesting ratios $\kappa_2/\kappa_1$.  
In each frame, we also show the three individual contributions, appropriately weighted with
products of $\kappa_i$ factors: $f_{11}$ (dashed blue),
$f_{22}$ (dot-dashed orange) and the interference term $f_{12}$ (dotted green).
The top left frame represents the case of the SM with $\kappa_1=1$ and $\kappa_2=0$.
The $M_{Z_2}$ distribution is comprised entirely of the $f_{11}$ component and peaks rather
sharply around $28$ GeV. As we start increasing the value of $\kappa_2$, the 
(negative) interference term $f_{12}$ begins to partially offset the 
$f_{11}$ piece and shifts the peak towards lower $M_{Z_2}$ values.
At the same time, the shape of the $M_{Z_2}$ distribution becomes deformed, 
while the $M_{Z_2}$ peak becomes rather broad.

\begin{figure}[t]
\hspace*{6pt}
\includegraphics[width=\2]{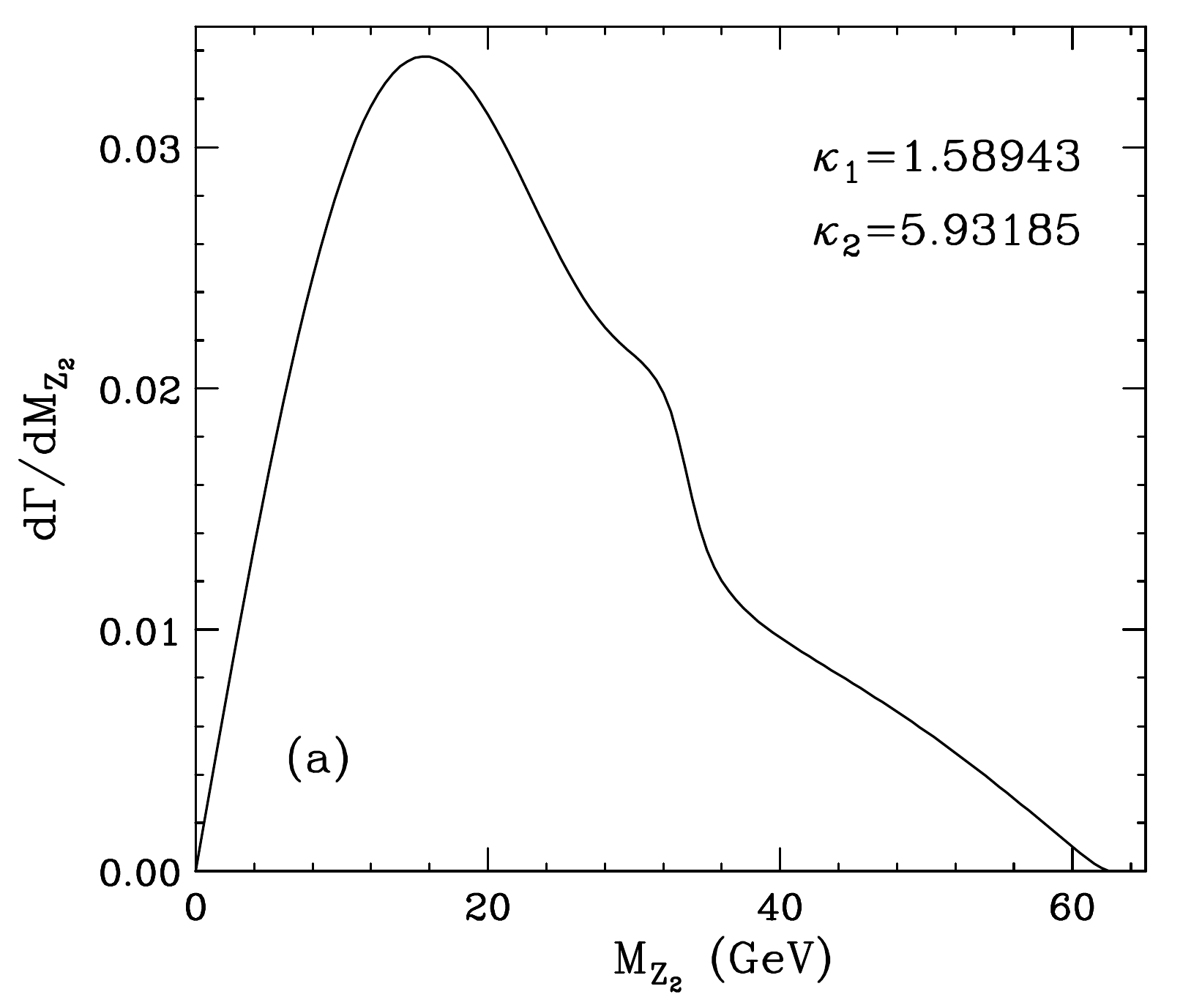} 
\includegraphics[width=\2]{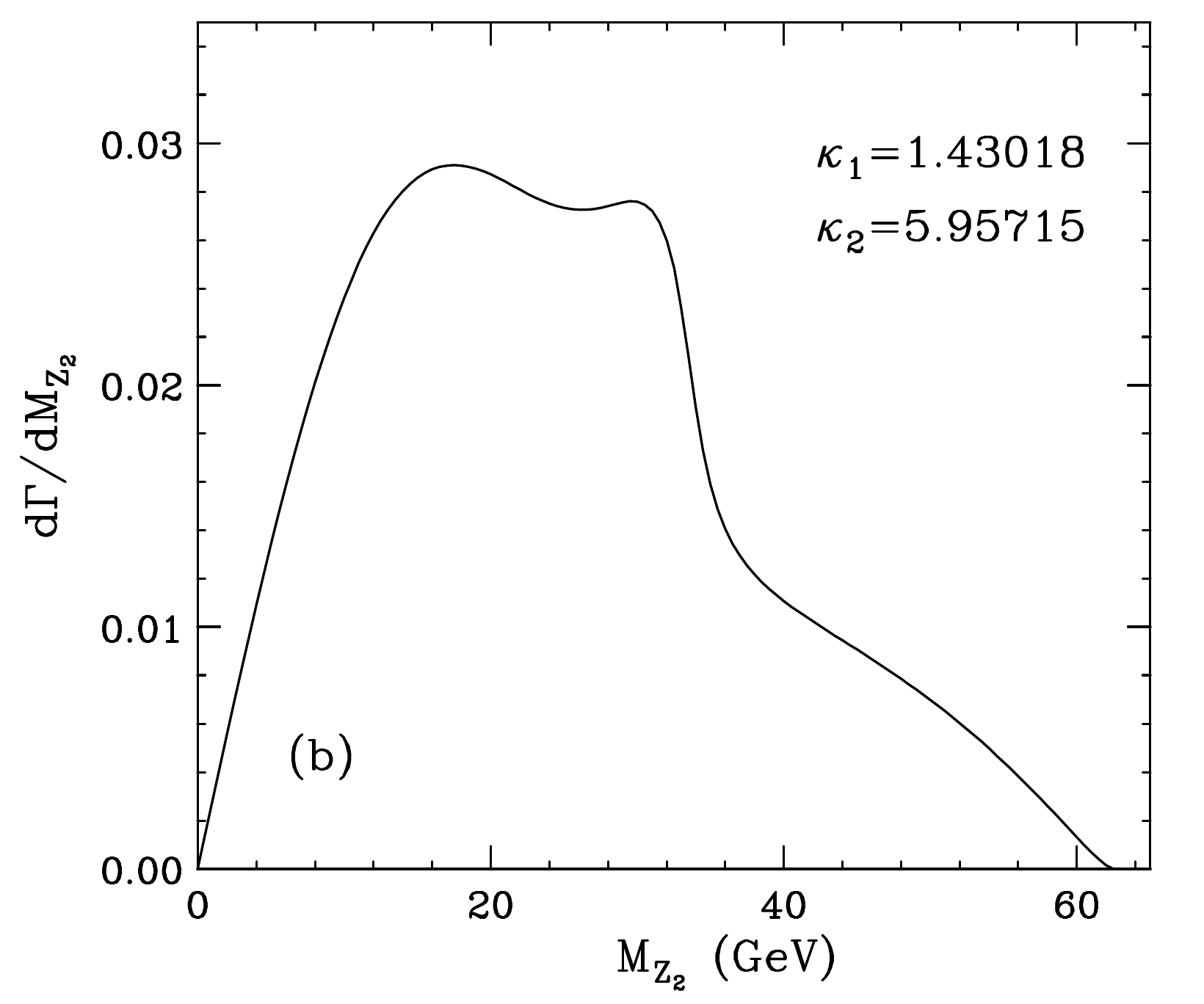} 
\includegraphics[width=\2]{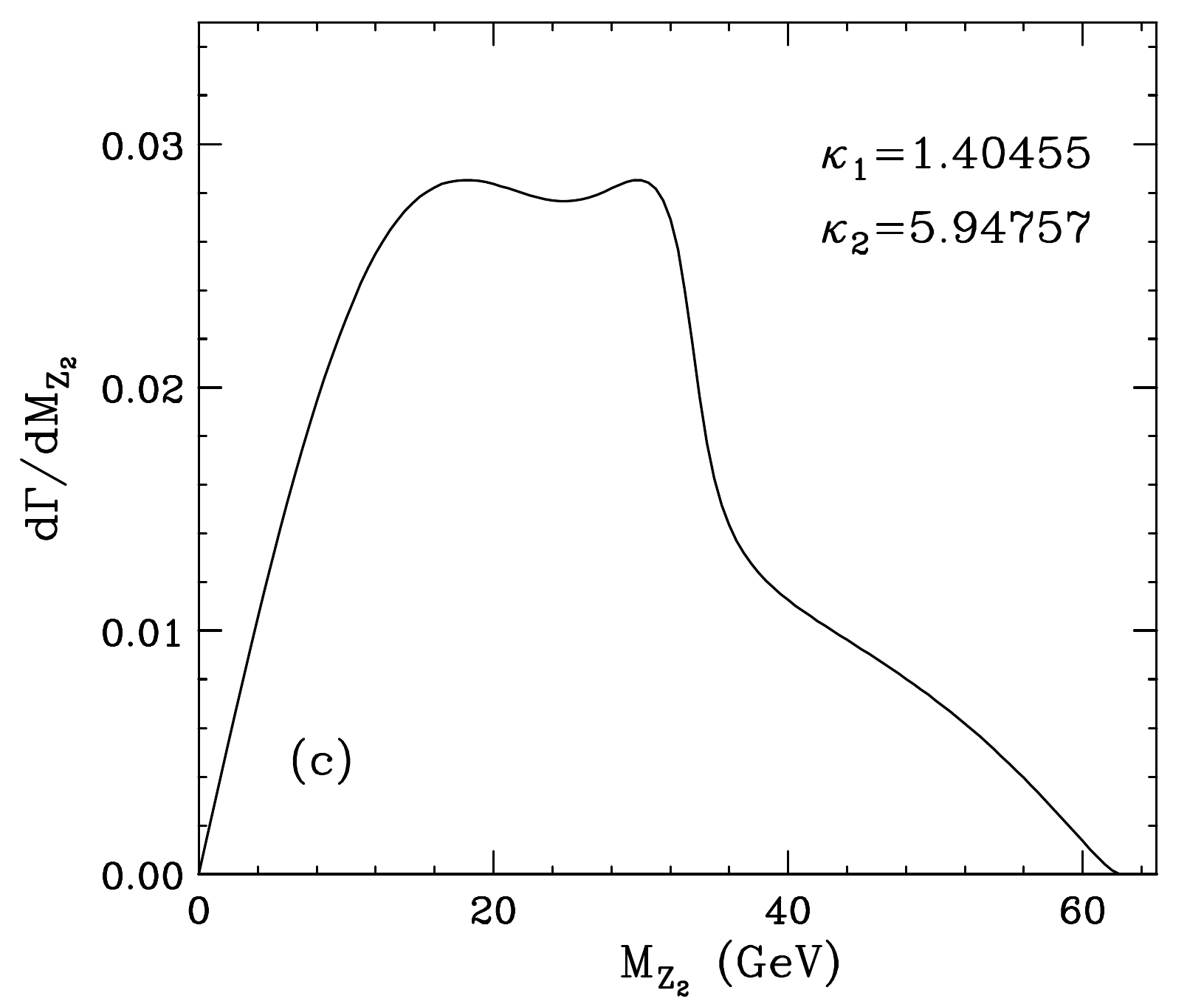} 
\includegraphics[width=\2]{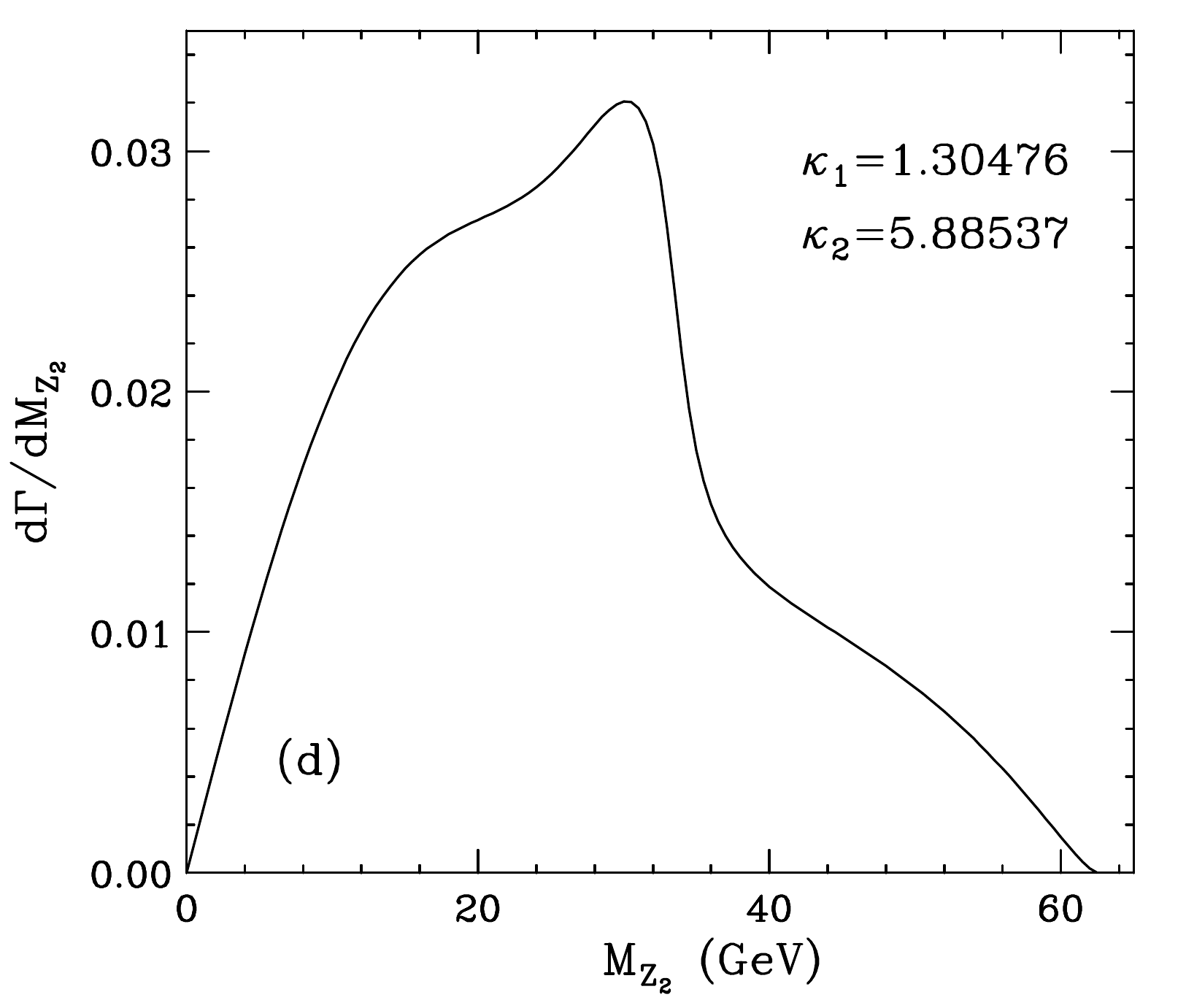} 
\caption{Unit-normalized differential $M_{Z_2}$ distributions for $\kappa_3=0$ and several choices of
$\kappa_1$ and $\kappa_2$ near the point of the ``first order phase transition", 
where the impact of interference 
on the shape of the distribution is maximal (see Fig.~\ref{fig:MZ2movie}). \label{fig:zoom}}
\end{figure} 

A very interesting situation occurs in the $\kappa_i$ parameter region 
illustrated by the plots in the second row of Fig.~\ref{fig:MZ2movie}.
Here the cancellation between the (negative) interference term $f_{12}$
and the (positive) $f_{11}$ and $f_{22}$ is near maximal
(see Eq.~(\ref{maxinterference})). More importantly, the resulting 
$M_{Z_2}$ distribution begins to develop a second local peak 
at high values of $M_{Z_2}\sim 30$ GeV. As $\kappa_2$ grows, this
secondary peak becomes stronger and eventually takes over as 
the primary peak in the distribution, causing the sudden jump
seen in Fig.~\ref{fig:peak}(a). This phenomenon resembles a 
``first order phase transition" and can be seen more clearly in
Fig.~\ref{fig:zoom}, where we zoom in on the actual $M_{Z_2}$ 
distribution without the individual contributions.
Of course, in the regime where this interesting behavior occurs, the
large destructive interference also suppresses the cross section.  The
reader will note that the values of $\kappa_1$ and $\kappa_2$ shown in
Fig.~\ref{fig:zoom}, which are necessary to give the correct SM partial 
width in Eq.~(\ref{GammaZZ}), are relatively large as a result.

As the value of $\kappa_2$ is increased beyond the region of the ``first order phase transition"
shown in Fig.~\ref{fig:zoom},
the $M_{Z_2}$ distribution starts to be dominated by the $f_{22}$ contribution
and eventually we get to the pure $0^+_{\mathrm{h}}$ state 
(the second to last panel in Fig.~\ref{fig:MZ2movie}). 
The final panel in Fig.~\ref{fig:MZ2movie} shows a representative point with 
opposite signs for the couplings $\kappa_1$ and $\kappa_2$. 
In that case, the sign of the interference term $f_{12}$ is flipped and it adds
constructively with $f_{11}$ and $f_{22}$, causing the peak of the $M_{Z_2}$
distribution to stay in the vicinity of $28-30$ GeV.

\section{Methodology}
\label{sec:method}

\subsection{Optimized Analyses}

To obtain the greatest sensitivity to a signal in a model which is characterized 
by a modest number of parameters, it is customary to use analyses 
with criteria specifically optimized for each point in the parameter 
space of the underlying model. This procedure is used in all
searches for Higgs bosons, whether SM or otherwise.
The approach has also been advocated for SUSY searches where the
signal model may have a greater number of
parameters~\cite{Matchev:1999nb,Matchev:1999yn}.
In line with this idea, we introduce the kinematic discriminants that are automatically optimized
for each point in the $XZZ$ coupling parameter space.

In this report we assume that the cross section has been well-measured and that
variations in the overall rate may be absorbed into the $ggX$
couplings (provided we consider only the  $pp \to X \to ZZ \to 4\ell$
channel). For this reason, the parameters we aim to measure 
are not the $XZZ$ couplings, $\kappa_i$, 
but their ratios $\kappa_2/\kappa_1$ and $\kappa_3/\kappa_1$. 
These quantities can be easily re-expressed in any desired convention,
such as ``geolocating'' angles as in Ref.~\cite{Gainer:2013rxa} or 
$f_{a}$-like fractions as in Ref.~\cite{CMS-PAS-HIG-13-002}.
In this study we assume that the couplings are real numbers, 
as already explained in Sec.~\ref{Subsec:loop}.

\subsection{Preparation of Monte Carlo Samples}

The analyses are performed using simulated $gg \to X \to ZZ \to 4\ell$ events,
generated using {\sc FeynRules}~\cite{FeynRules} and {\sc
  MadGraph}~\cite{MadGraph} according to the {\sc MEKD} framework~\cite{MEKD}.
This approach ensures that we include all interference effects:
those arising from the presence of multiple terms in the Lagrangian 
as well as those associated with permutations of identical leptons
in the $4e$ and $4\mu$ final states. 
Following the ATLAS and CMS results~\cite{ATLAS-CONF-2013-013, CMS-PAS-HIG-13-002} the mass of the scalar Higgs-like boson mass is taken to be $125$~GeV.
We use {\sc MadGraph} to simulate the $q\bar{q} \to ZZ$
backgrounds.
Our simulation is performed entirely at the leading order and at the parton level.  
In order to compensate somewhat, we consider events with the four-lepton invariant mass
 in a very conservative $10$ GeV mass window centered at the Higgs mass of $125$ GeV 
(in contrast, the LHC detectors have $1-2\%$ mass resolution). 
Consequently, the larger mass window results in the
acceptance of more background events.

We use lepton kinematic selection criteria very similar to those used in the $H \to ZZ \to 4\ell$ analyses of ATLAS
and CMS experiments~\cite{ATLAS-CONF-2013-013,
  CMS-PAS-HIG-13-002}.
Leptons are required to have transverse momenta $p_T > 5$~GeV and
pseudorapidity $|\eta| < 2.5$. At least one same-flavor opposite-sign
lepton pair must have an invariant mass greater than $40$~GeV, 
while the other lepton pair must have an invariant mass greater than
$12$~GeV.
We use events with all three final-state combinations ($4e$, $4\mu$,
and $2e2\mu + 2\mu2e$) in all of our analyses.

\subsection{Projected Event Yields}

In order to obtain the analysis results as a function of the integrated luminosity
of the LHC runs at $14$ TeV, 
we estimate experimental reconstruction efficiencies and contribution of the background at the $14$ TeV
LHC using the average of the expected signal and background event yields 
reported by ATLAS and CMS (Table~\ref{tab:EventYields}). 
The number of events expected in the $14$ TeV LHC runs with $L$ fb$^{-1}$ of integrated luminosity, $N(L)$, is computed as:
\begin{equation} 
\label{eq:ScalingRule}
N(L) = \frac { N_{\mathrm{ATLAS}} + N_{\mathrm{CMS}} } {2} \times 
       \frac { \sigma(14~\mathrm{TeV}) } { \sigma(8~\mathrm{TeV}) } \times
       \frac { L } { (25~\mathrm{fb}^{-1}) },
\end{equation}
The ratios of cross sections for the SM Higgs boson signal and the
dominant $q\bar{q} \to ZZ$ background used in Eq.~(\ref{eq:ScalingRule})
are $\sigma_H(14~\mathrm{TeV})/\sigma_H(8~\mathrm{TeV}) =
2.6$~\cite{Dittmaier:2011ti} 
and $\sigma_{ZZ}(14~\mathrm{TeV})/\sigma_{ZZ}(8\mathrm{TeV}) = 1.9$
(computed with {\sc MCFM}~\cite{MCFM}). 
With these assumptions, the average expected event rates per
experiment per fb$^{-1}$ of integrated luminosity at the $14$ TeV LHC
are $1.9$ (signal) and $0.76$ (background, in the $10$ GeV mass window
described above).

\begin{table*}[th]
\begin{center}
\small
  \caption[ ] {The expected event yields for the SM Higgs boson signal with mass $m_H=125$~GeV and background, as reported by the ATLAS and CMS collaborations for 7+8~TeV LHC Run I.}
  \label{tab:EventYields}
\begin{tabular}{ l c  c  c  c  }
\hline 
\hline 
Experiment                    & Process    & Event yield             & Integrated luminosity at 7 + 8 TeV & Source \\ 
\hline 
\multirow{2}{*}{ATLAS}        & Signal     &     18.2                & \multirow{2}{*}{ 4.6 + 20.7 = 25.3~fb$^{-1}$}      &  Tab.~7 in Ref.~\cite{ATLAS-CONF-2013-013} \\
                              & Bkgd       &     $\sim$1 event/GeV   &                                                    &  Fig.~4 in Ref.~\cite{ATLAS-CONF-2013-013} \\
\hline 
\multirow{2}{*}{CMS}          & Signal     &     19.2                & \multirow{2}{*}{ 5.1 + 19.6 = 24.7~fb$^{-1}$}      &  Tab.~2  in Ref.~\cite{CMS-PAS-HIG-13-002} \\
                              & Bkgd       &     $\sim$1 event/GeV   &                                                    &  Fig.~2 in Ref.~\cite{CMS-PAS-HIG-13-002} \\
\hline 
\hline 
\end{tabular}
\end{center}
\end{table*}

\subsection{Kinematic Discriminants}

Kinematic discriminants for separation between the two types of four-lepton processes, $A$ and $B$, may be constructed by calculating the ratio of the squared matrix elements
for these two hypotheses, as described in Ref.~\cite{MEKD}.
For each four-lepton event with kinematic information ${\bf{x}}$, one can compute:
\begin{equation}
\label{eq:KD}
D(A,B; \mathbf{x}) =  \frac {|\mathcal{M}(A;\mathbf{x})|^2}
{|\mathcal{M}(B;\mathbf{x})|^2}.
\end{equation}

In our analysis, we compute the kinematic discriminants following this approach.
We first consider the kinematic discriminant $D(X;0^+)$.  Here, the
hypothesis ``$X$'' is the hypothesis that the scalar
Higgs-like boson couples to $Z$s via both the $\kappa_1$ and
$\kappa_3$ operators. We will further refer symbolically to this state as $X = \kappa_1 \, [0^+] + \kappa_3 \, [0^-]$.
The hypothesis ``$0^+$'' assumes that the scalar Higgs-like boson has
only the tree-level SM coupling to $Z$ bosons. Therefore, for $D(X;0^+)$ we obtain:
\begin{equation}
\label{eq:D(X;H)}
D(X;0^+) = \frac {|\mathcal{M}(X)|^2} {|\mathcal{M}(0^+)|^2} 
         = \kappa_1^2 + 
           \kappa_3^2 \frac{|\mathcal{M}(0^-)|^2}{|\mathcal{M}(0^+)|^2}  + 
           \kappa_1 \kappa_3 \frac{(\mathrm{interference})}{|\mathcal{M}(0^+)|^2}.
\end{equation}
By construction, this discriminant takes into account all aspects in
which kinematic distributions differ between the two hypotheses, including
in particular those associated with the interference between the
$\kappa_1$ and $\kappa_3$ operators in hypothesis ``$X$''.  

Alternatively, one can choose to use the kinematic discriminant
$D(0^-;0^+)$~\cite{CMS-PAS-HIG-13-002}, 
where the two hypotheses, ``$0^-$'' and ``$0^+$'', correspond to the cases where
only the $\kappa_3$ term or only the $\kappa_1$ term are non-vanishing, respectively:
\begin{equation}
\label{eq:D(A;H)}
D(0^-;0^+) = \frac {|\mathcal{M}(0^-)|^2} {|\mathcal{M}(0^+)|^2}.
\end{equation}
Since the two hypotheses from which the discriminant is
calculated correspond to two pure states, discriminant $D(0^-;0^+)$ is explicitly \emph{insensitive} to
the potential effects on kinematic distributions associated with the
interference (unlike discriminant $D(X;0^+)$).
The $D(0^-;0^+)$ discriminant is optimal for comparing the two pure states or
for testing for the presence of an additional pseudoscalar state nearly degenerate with the
scalar Higgs-like boson (but with a sufficiently different mass that
there is no significant interference in the scalar and pseudoscalar
production and decays).  However, as it ignores interference effects, it
is not optimal for measuring the state $X$ which couples with $ZZ$
via both $\kappa_1$- and $\kappa_3$- terms.
Discriminant $D(X;0^+)$ described above 
is ideal for this purpose.

\subsection{Statistical analysis}

\begin{figure}[h]
\subfigure[]{\includegraphics[width=\2]{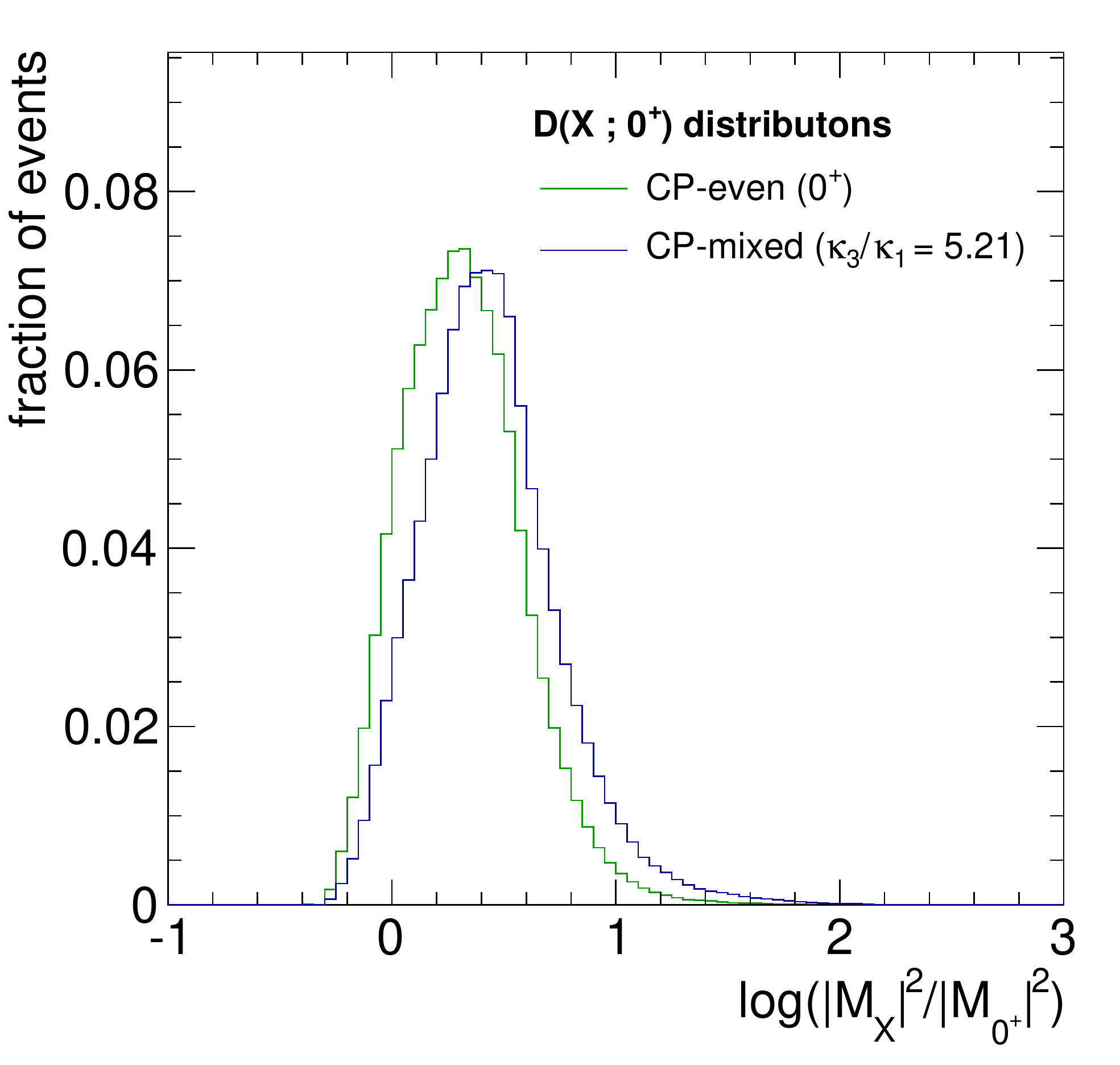} 
\label{fig:D_distributions}}
\hspace*{6pt}
\subfigure[]{\includegraphics[width=\2]{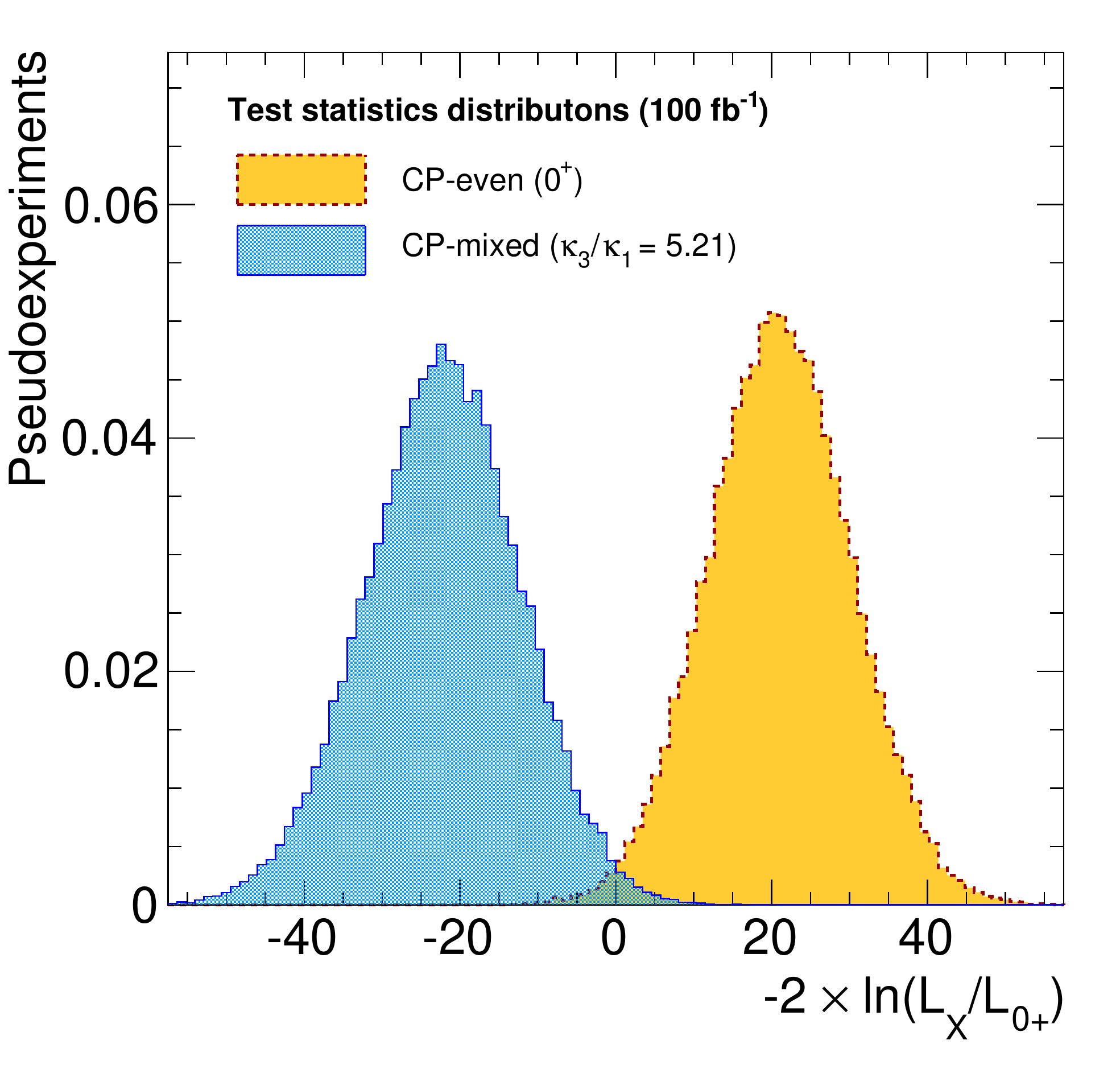} 
\label{fig:TestStat_distributions}}
\caption{
(a) Distributions of $D(X;0^+) = \frac {|\mathcal{M}(X)|^2} {|\mathcal{M}(0^+)|^2}$
for two alternative hypotheses ``$0^+$'' and ``$X$'', where $X$ has $\kappa_3/\kappa_1=5.21$. 
(b) Corresponding distributions for the test statistic defined in
Eq.~(\ref{eq:q}) for pseudoexperiments 
at an integrated luminosity $L = 100$~fb$^{-1}$.}
\end{figure}

We obtain distributions for the
kinematic discriminants described above using simulation. 
Distributions are obtained for events that correspond to the signal hypothesis ``$X$'',
to the signal hypothesis ``$0^+$'' (both described above) and to the background hypothesis.
Examples of the distributions, $pdf(D \, | \, X + \mathrm{bkg})$
and $pdf(D \, | \, 0^+ + \mathrm{bkg})$ are shown in Fig.~\ref{fig:D_distributions}.

These kinematic discriminant distributions are then used to construct
the test statistic $q$ as follows:
\begin{equation}
\label{eq:q}
q = -2 \, \ln \frac {\mathcal{L}(\mathrm{``data"} \, | \, X + \mathrm{bkg})}
					 {\mathcal{L}(\mathrm{``data"} \, | \, 0^+ + \mathrm{bkg})}=
    -2 \, \ln \prod_{i} \frac {pdf(D_i \, | \, X + \mathrm{bkg})} {pdf(D_i \, | \, 0^+ + \mathrm{bkg})},
\end{equation}
where $i$ runs over all the events in an pseudoexperiment.
An example of the test statistic distributions obtained with $50000$ pseudoexperiments for a particular choice of the integrated luminosity $L$ and $\kappa_3/\kappa_1$  ratio is shown in
Fig.~\ref{fig:TestStat_distributions}.
 
To quantify the expected separation power between alternative signal hypotheses,
we find a ``mid-point'' value,  $\tilde q$, of the test statistic $q$
between the medians of the two test statistic distributions (those
generated using each signal hypothesis). We use point $\tilde q$ to define 
two ``tail probabilities'', $P(q \geq \tilde q \, | \, X)$ and $P(q
\leq \tilde q \, | \, 0^+)$,  in such a way that
$P(q \geq \tilde q \, | \, X) = P(q\leq \tilde q \, | \, 0^+)$.
This tail probability is then converted into significance $\tilde Z$ (in $\sigma$) 
using the one-sided Gaussian tail convention:
\begin{equation}
\label{eq:Z}
P \, = \, \int_{\tilde Z}^{+\infty} \frac{1}{\sqrt{2\pi}} \exp(-x^2/2) \,\, \mathrm{d}x.
\end{equation}
Finally, for the separation power between alternative signal hypotheses
we quote $Z = 2 \tilde Z$, where the extra factor of 2 arises from
the fact that the $\tilde q$ point is half-way between the medians of
the two distributions. With such a definition, we treat two alternative hypotheses 
symmetrically and we do not need to generate billions of pseudoexperiments
to assess tail probabilities corresponding to $5 \sigma$-separations.

The presence of a non-zero value of $\kappa_3$ could be established,
albeit with different significances, in searches performed using
either $D(X;0^+)$ or $D(0^-;0^+)$.
The difference in the sensitivity between the two searches is manifested 
in case the interference between the $\kappa_1$ and $\kappa_3$ operators is
present.
This is not unlike the actual discovery of the Higgs boson candidate,
which gave rise to the $\sim$$5\sigma$ signal in the SM Higgs
search~\cite{Aad:2012tfa, Chatrchyan:2012ufa} and at the same time was
also seen as $\sim$$3\sigma$ excesses in the Higgs boson searches 
performed in the context of the fermiophobic and SM4
scenarios~\cite{Chatrchyan:2013sfs}. In case the presence of a non-zero value of $\kappa_3$ is
established,
 the next two questions to answer are: 
\begin{itemize}
\item whether there is one state $X = \kappa_1 \, [0^+] + \kappa_3 \, [0^-]$ with interference 
or there are two non-interfering states, scalar $S = \kappa_1
\, [0^+]$ and pseudoscalar $P = \kappa_3 \, [0^-]$; 
\item if there is interference, how well we can tell apart the
  relative signs of $\kappa_3$ and $\kappa_1$ couplings. 
\end{itemize}
%
\begin{figure}[h]
\includegraphics[width=\1]{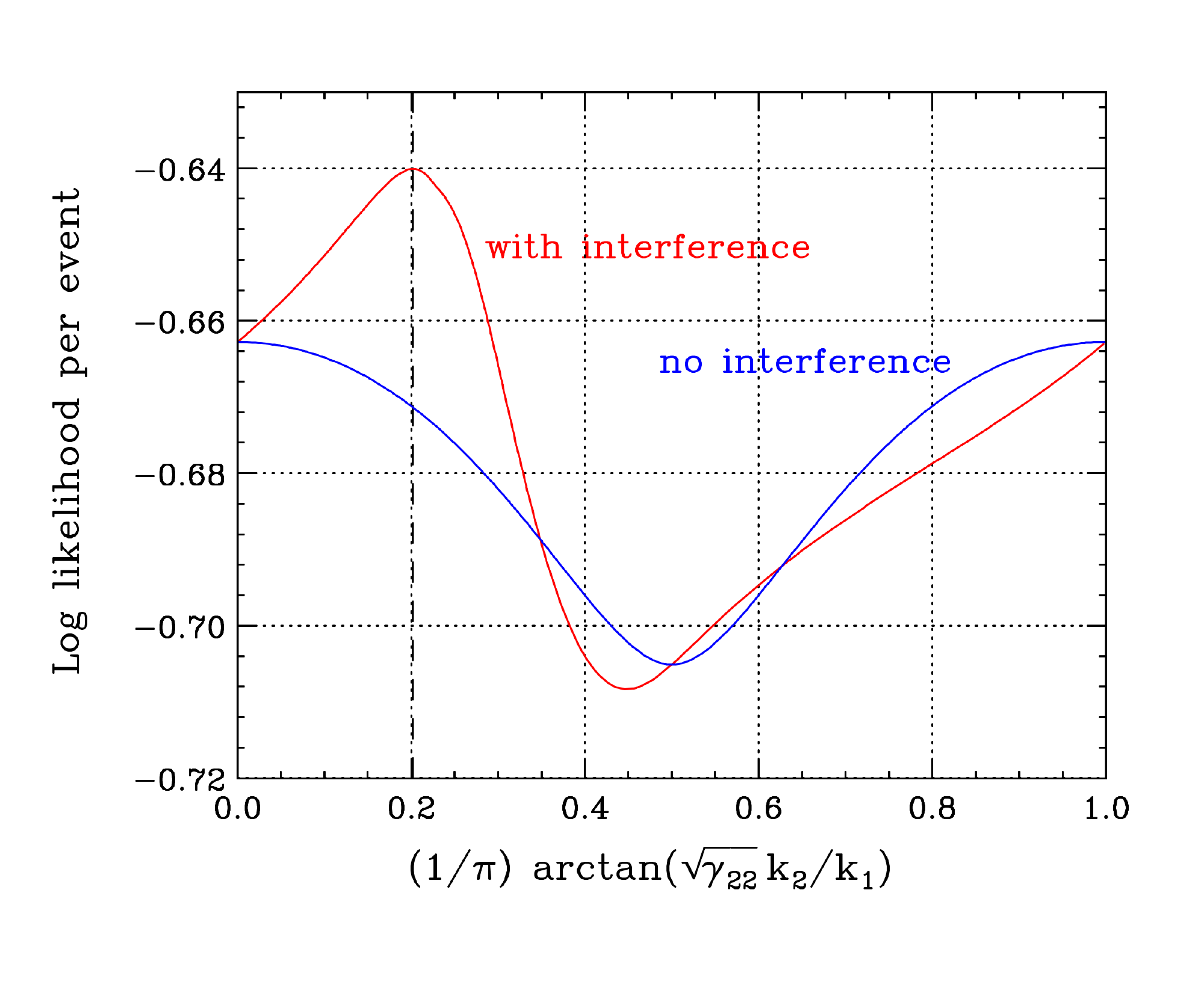}
\caption{
The log likelihood per event for various values of
$\kappa_2/\kappa_1$ for a particular benchmark point with $(\kappa_1,
\kappa_2) \approx (1.77,4.26)$ (vertical line), which is the point with the
same cross section as SM for which $x_1 = x_2$ in the
language of Ref.~\cite{Gainer:2013rxa}.  The quantity on the horizontal
axis represents the angle along a circle of constant cross section in
$(\kappa_1, \kappa_2)$ space in the absence of interference.}
\label{fig:likelihood_k1k2}
\end{figure} 
\begin{figure}[h]
\includegraphics[width=\1]{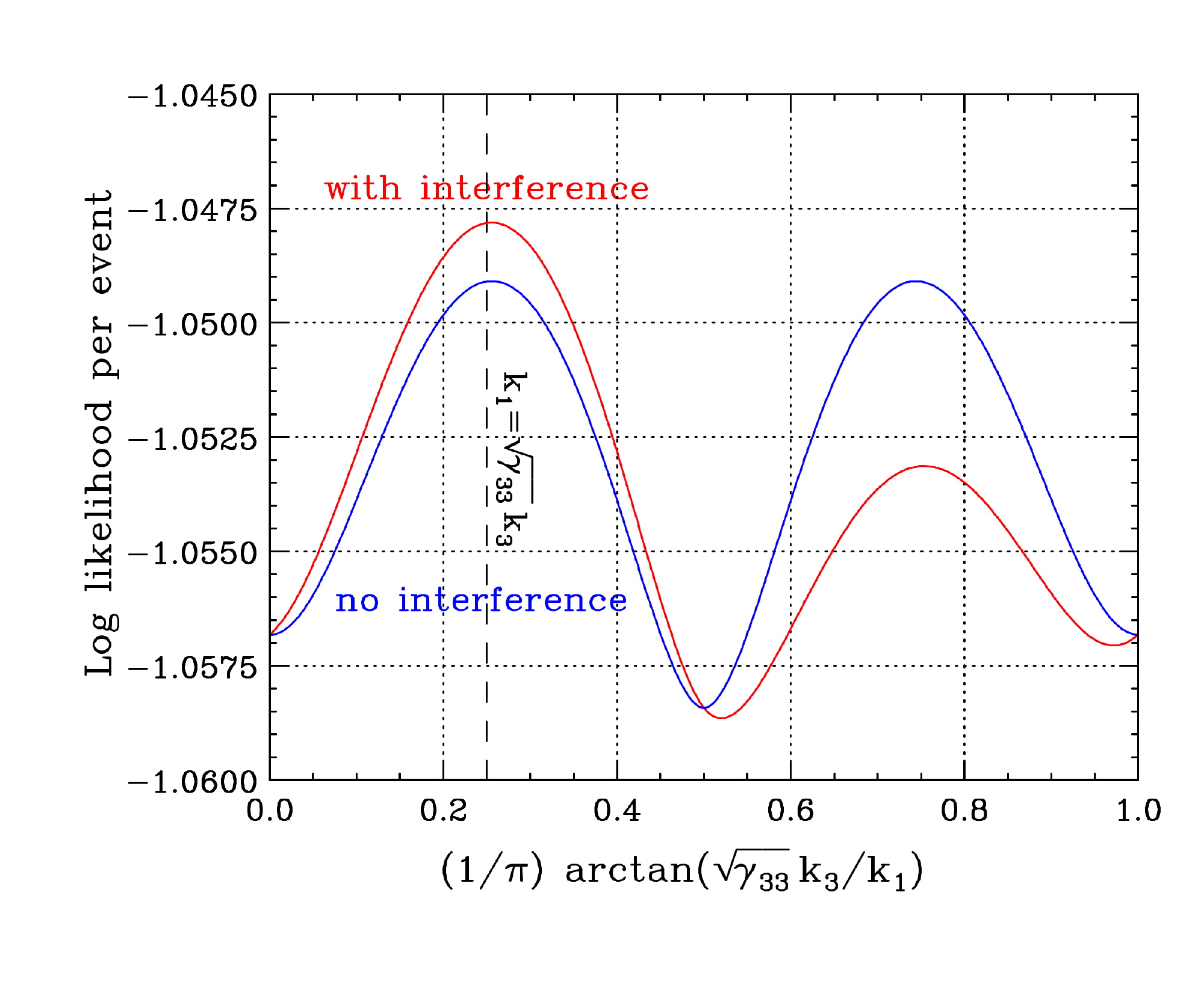} 
\caption{
The log likelihood per event for various values of
$\kappa_3/\kappa_1$ for a particular benchmark point with $(\kappa_1,
\kappa_3) = (1/\sqrt{2}, 1/\sqrt{2 \gamma_{33}})$ (vertical line),
which is the point with the same cross section as the standard model
and an angle of $\pi/4$ with respect to the SM axis, along a circle
of constant cross section in $(\kappa_1, \kappa_3)$ space.
}
\label{fig:likelihood_k1k3}
\end{figure} 
Both of these questions can be addressed by repeating the statistical analysis
with properly adjusted kinematic discriminants.  To demonstrate the
ability of an experiment to establish the presence or absence of interference,
as well as to determine the relative sign of couplings,
we plot the per event log likelihood for two particular benchmark
points in Figures~\ref{fig:likelihood_k1k2}
and~\ref{fig:likelihood_k1k3}.  The benchmark point used for
Figure~\ref{fig:likelihood_k1k2} (Figure~\ref{fig:likelihood_k1k3}) has
non-zero values for $\kappa_1$ and $\kappa_2$ ($\kappa_1$ and
$\kappa_3$), while the log likelihood is evaluated for various values of the $\kappa_2/\kappa_1
(\kappa_3/\kappa_1)$ ratio.

In the absence of interference, the likelihood
functions are symmetric under $\kappa_{2,3} \to - \kappa_{2,3}$.
The presence of interference breaks this symmetry and gives
one sensitivity to the sign of the couplings.  We note that
interference between contributions to the amplitude from the
$\kappa_1$ and the $\kappa_2$ terms is relatively
straightforward to detect, as one would expect from the behavior of
the $M_{Z2}$ distribution discussed above. On the other side, interference involving
the $\kappa_1$ and $\kappa_3$ terms will be more challenging to detect.
Interestingly, it is significantly easier to determine the correct sign
of $\kappa_3$ assuming interference, than it is to determine whether that
interference is present.

\section{Results}
\label{sec:results}

\begin{figure}[h]
\subfigure[]{\includegraphics[width=\2]{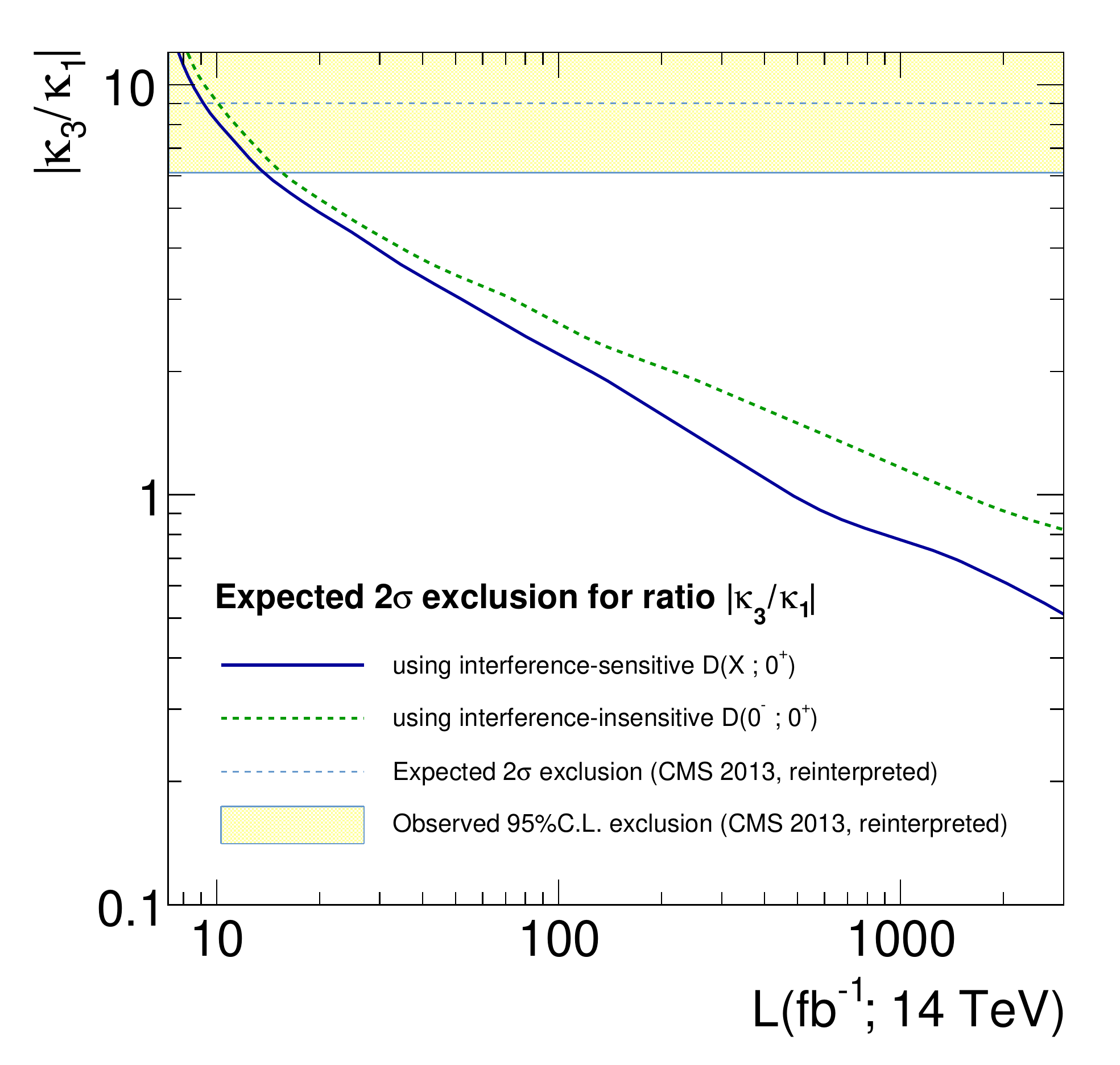} 
\label{fig:LimitsSensitivity}}
\hspace*{6pt}
\subfigure[]{\includegraphics[width=\2]{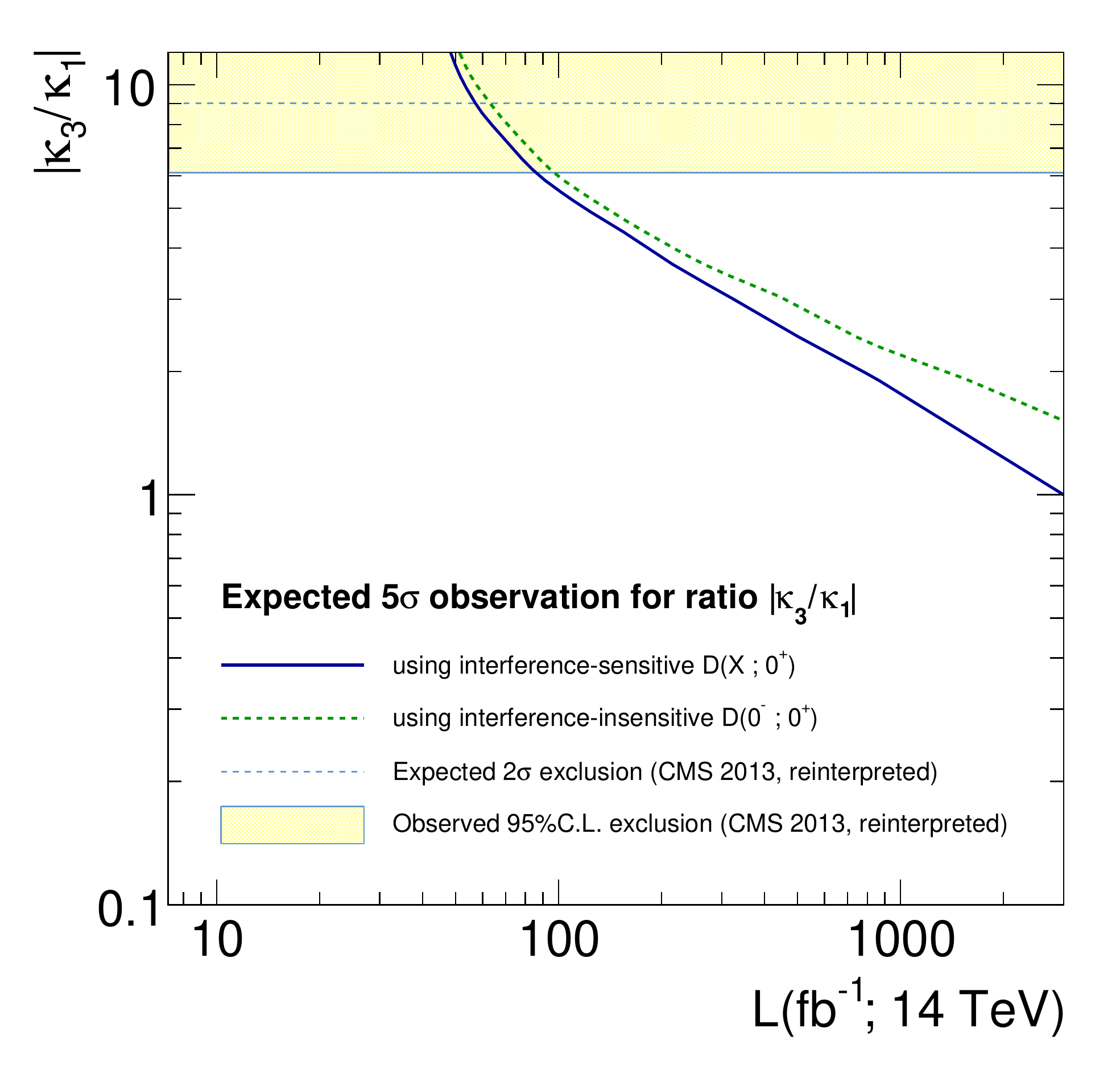} 
\label{fig:ObservationSensitivity}}
\caption{(a) The integrated luminosity required for an expected
  $2\,\sigma$-exclusion of the ratio of couplings $\kappa_3 /
  \kappa_1$, provided the data is described by the SM hypothesis. (b)
  The integrated luminosity required for $5\,\sigma$-observation
  sensitivity for the ratio of couplings $\kappa_3 / \kappa_1$, in the
  presence of non-zero $\kappa_3$. 
Results with interference-sensitive $D(X;0^+)$ and 
interference-blind $D(0^-;0^+)$ discriminants 
are shown with blue and green curves, respectively. }
\end{figure}

Figure~\ref{fig:LimitsSensitivity} presents the expected upper limits on 
the ratio of couplings $\kappa_3 / \kappa_1$ versus the integrated luminosity.
Similarly, Fig.~\ref{fig:ObservationSensitivity} shows a plot for the expected $5\,\sigma$-observation sensitivity. 
Results with both the optimal $D(X;0^+)$ and the interference-blind $D(0^-;0^+)$ 
discriminants are shown. 
The expected exclusion and observation sensitivities are identical for positive and
negative signs of the $\kappa_3 / \kappa_1$ ratio.  Of course, for a
given pseudoexperiment and in the actual LHC running one sign or the other
will be preferred by the data.

In Fig.~\ref{fig:LimitsSensitivity} and Fig.~\ref{fig:ObservationSensitivity} one can see that the sensitivities obtained with the two discriminants scale very differently 
with integrated luminosity $L$. 
This is because the $D(0^-;0^+)$ discriminant does not change when one wishes to probe
smaller or larger values of the $\kappa_3 / \kappa_1$ ratio. 
In this case, the sensitivity 
to $|\kappa_3 / \kappa_1|^2$ which is related to the ratio of cross
sections $\sigma_3 / \sigma_1 = \gamma_{33} \kappa_3^2/ \kappa_1^2$
scales approximately as $1/\sqrt{L}$. 
On the other hand, the $D(X;0^+)$ discriminant is automatically optimized 
for any given $\kappa_3 / \kappa_1$-value probed.
For this reason analyses with $D(X;0^+)$
which probe different fractions of the $0^-$ state can be thought of as
separate analyses, and their respective sensitivities to $|\kappa_3 /
\kappa_1|^2$ at different luminosities do not have to be connected via
a simple $1/\sqrt{L}$ relationship.

The difference between the sensitivities obtained 
with the two discriminants can be quantified in terms of
a ratio of integrated luminosities required to achieve the same
sensitivity. Figures~\ref{fig:LimitsSensitivity} and \ref{fig:ObservationSensitivity} show that
this difference grows very large for smaller values of $\kappa_3 / \kappa_1$.
For example, to probe $\kappa_3 / \kappa_1 = 1$, the
integrated luminosities needed for a $2\sigma$-separation differ by a factor of $4$:
$\sim 700$\fb\ with the interference-sensitive $D(X;0^+)$ discriminant 
versus $\sim 3000$\fb\ with the interference-blind discriminant $D(0^-;0^+)$.
With $L=3000$\fb, the interference-sensitive discriminant $D(X;0^+)$ allows for reaching a 
$5\sigma$-sensitivity for $|\kappa_3/\kappa_1| \sim 1$.
However, at an integrated luminosity of 10\fb which approximately corresponds 
to 25\fb\ at 8~TeV, the difference in sensitivities to $\kappa_3 / \kappa_1$
achievable with the two discriminants is rather modest, $\mathcal{O}(10$\%).

\begin{figure}[h]
\subfigure[]{\includegraphics[width=\2]{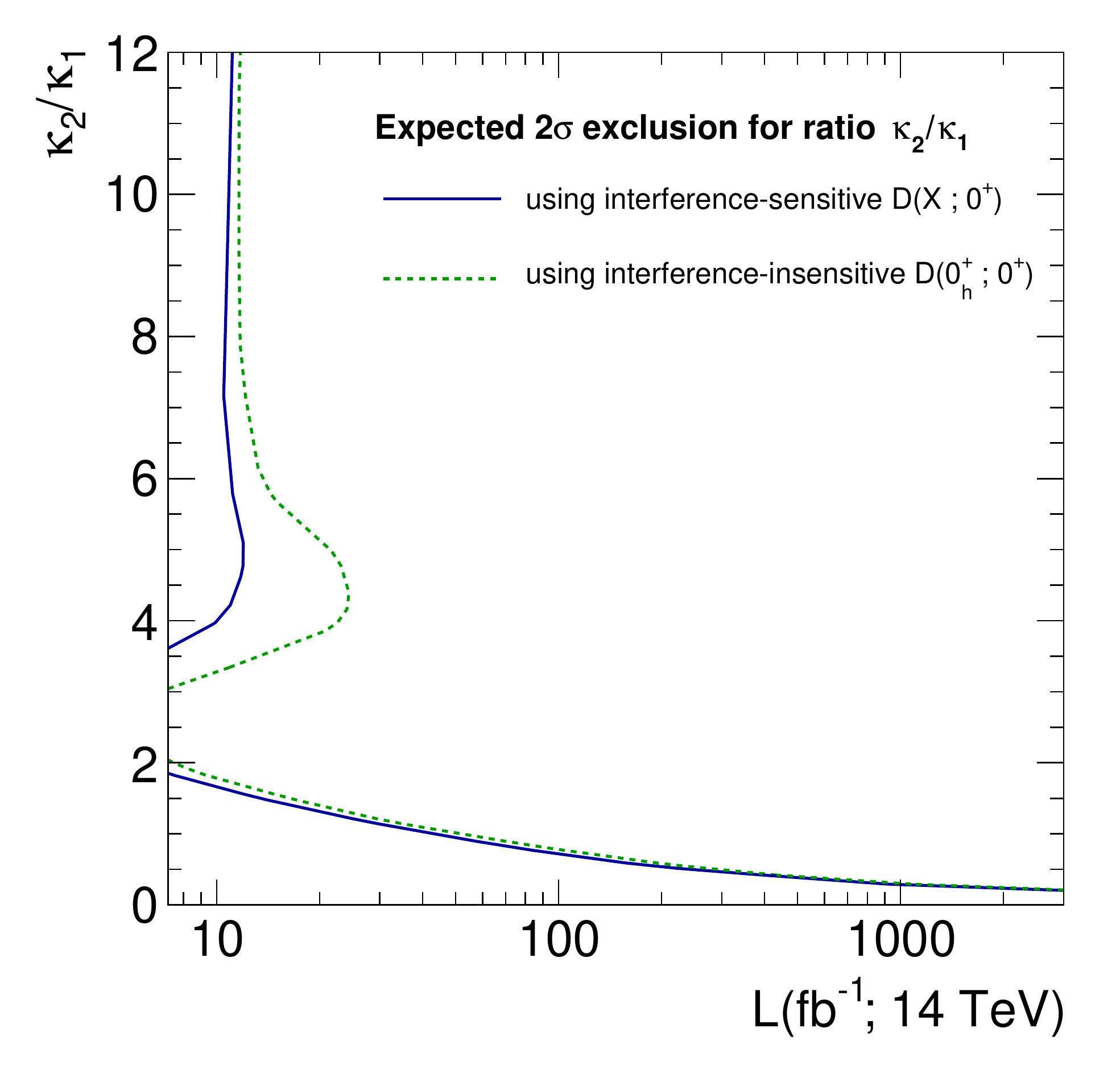}}
\hspace*{6pt}
\subfigure[]{\includegraphics[width=\2]{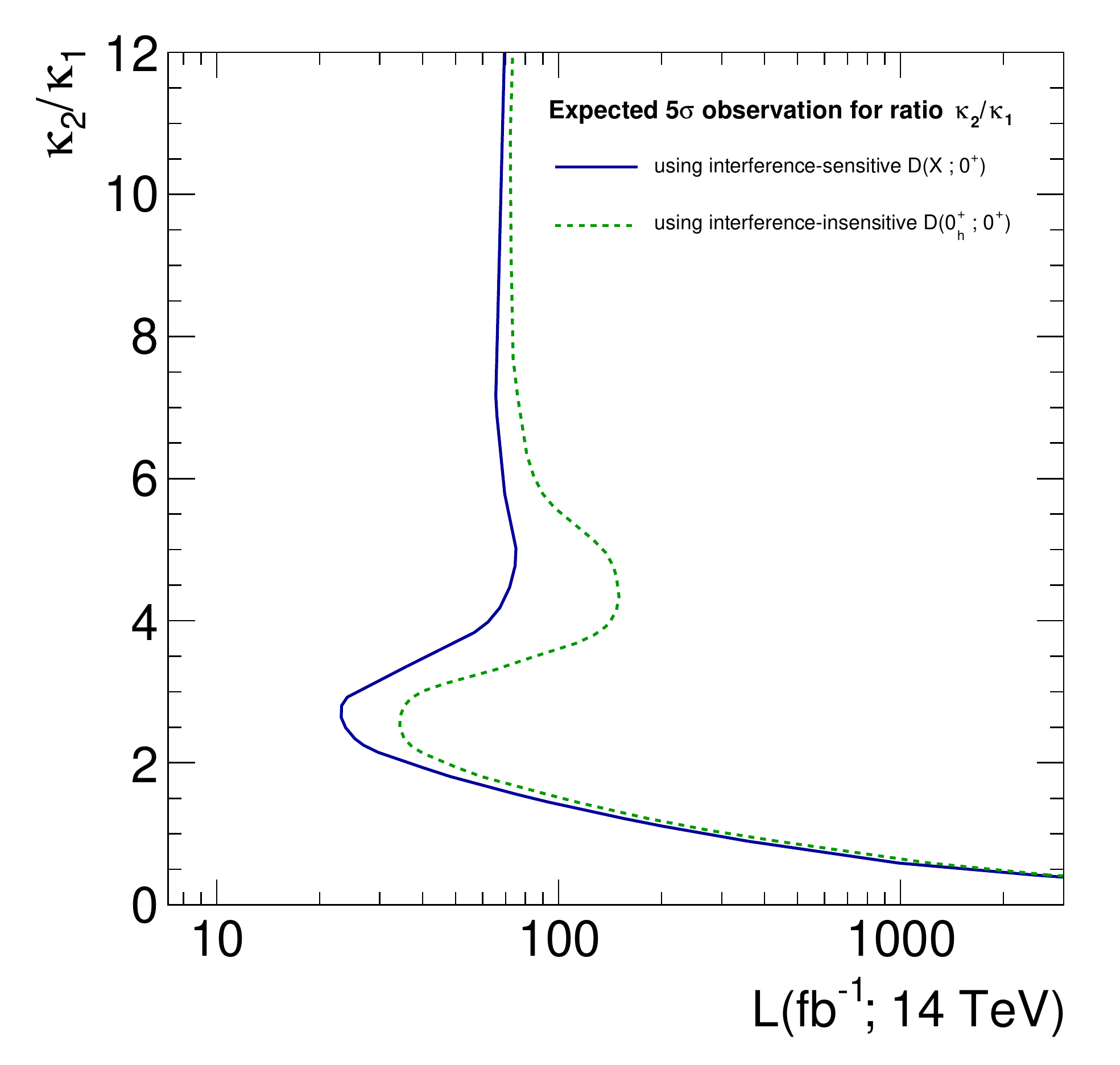}} 
\caption{(a) The integrated luminosity required for a 95\% CL
  exclusion of the ratio of couplings $\kappa_2 / \kappa_1$, 
  provided the data is described by the SM hypothesis.
Results with both $D(X;0^+)$, $X = \kappa_1 \, [0^+] + \kappa_2 \, [0_h^+]$, 
and interference-blind $D(0_h^+;0^+)$ discriminants are shown. 
(b) Luminosity required for a $5\,\sigma$-observation of a presence 
of a $\mathrm{J^{CP}}=0_h^-$ state versus assumed ratio of couplings $\kappa_2 / \kappa_1$.}
\label{fig:ZeroPlusSensitivity}
\end{figure}

Figure~\ref{fig:ZeroPlusSensitivity} shows the expected $2\,\sigma$-exclusion 
and $5\,\sigma$-observation sensitivities for 
the ratio of couplings $\kappa_2 / \kappa_1$ vs. the integrated
luminosity. In these figures we focus on the $\kappa_2/ \kappa_1 > 0$ region for
which destructive interference is present, as the prospects for early
detection are more favorable with this choice of the relative sign of the two couplings. 
Results with both the optimal $D(X;0^+)$, where $X = \kappa_1 \, [0^+] + \kappa_2 \, [0_{\mathrm{h}}^+]$, 
and interference-blind $D(0_{\mathrm{h}}^+;0^+)$ discriminants are shown. 
As suggested in Eq.~(\ref{maxinterference}) above, there is substantial destructive
interference in the range of $\kappa_2/\kappa_1 \approx 2-4$ that leads to dramatic changes
in the $M_{Z_2}$ invariant mass distribution shown in
Fig.~\ref{fig:MZ2movie}. The kinematic discriminants are
automatically sensitive to such changes in the $M_{Z_2}$ distributions, 
as well as to changes in other kinematic variables. As a results, it would
be relatively easy to differentiate the case where $\kappa_2/\kappa_1$
is in this range from the pure SM Higgs-like boson.
In fact, since the $\sim 25$\fb~of $8$ TeV data already recorded on tape translates
into the $\sim 10$\fb~of $14$ TeV data, we find that the LHC experiments
should already be able to discover or exclude the $2 < \kappa_2/\kappa_1 < 4$ range.
We note also that with existing data there should be a borderline sensitivity for exclusion
of the $\kappa_2/\kappa_1 > 4$ range, which includes the case of a
pure $0^+_{\mathrm{h}}$ state. 
This result is well in
agreement with the expected sensitivity of 1.8$\sigma$ for a 100\% pure $0^+_{\mathrm{h}}$ state
reported by CMS~\cite{CMS-PAS-HIG-13-002}. The observed limit reported by CMS is $92\%$~CL.

\section{Summary}
\label{sec:summary}

We have considered the important question of how to measure the
couplings of the scalar Higgs-like boson, $X$, to two $Z$ bosons.  In
particular, we have studied the effects of the interference between various
$XZZ$ operators, presented the kinematic discriminants that take into account
these interference effects, and provided projections for the coupling measurements
using these discriminants at the $14$ TeV LHC.
  
We have also compared the sensitivity of these kinematic discriminants with
the kinematic discriminants that do not include interference terms and
found that incorporating interference effects allows  
one to significantly improve the sensitivity to states where more than
one operator is present in the $XZZ$ coupling. 
Depending on the value of the couplings being probed,
using analyses that take interference into account
may reduce the integrated luminosity required to reach a given
sensitivity by as much as a factor of four, as compared with analyses
that neglect this interference.  Thus using analyses such as those
presented may allow one to reach given sensitivity benchmarks at the LHC 
years earlier than otherwise.

\section*{Acknowledgements}
\label{sec:acknowledgements}

We thank our CMS colleagues for useful discussions.
M.~Park is supported by the CERN-Korea fellowship through the National
Research Foundation of Korea.
Work supported in part by U.S.~Department of Energy Grant
DE-FG02-97ER41029 and NSF Grant 1007115.


\bibliography{SpinZeroTensorWithInterference}

\end{document}